\documentclass[aps,prb,twocolumn,longbibliography,superscriptaddress,floatfix,10pt]{revtex4-2}

\usepackage[utf8]{inputenc}
\usepackage{amssymb,amsfonts,amsmath,dsfont,amsmath}
\usepackage{graphicx}
\usepackage{textcomp}
\usepackage{gensymb}
\usepackage{pifont}
\usepackage{bm}
\usepackage{color}
\usepackage{accents}
\usepackage{mathtools}
\usepackage{nicefrac}
\usepackage{url}
\usepackage{multirow}
\usepackage{hyperref}
\usepackage[capitalize]{cleveref}
\usepackage{tabularx}
\usepackage{upgreek}
\usepackage{chemformula}
\hypersetup{colorlinks=true,citecolor=blue,urlcolor=blue,linkcolor=blue}
\usepackage{graphicx}
\usepackage[caption=false]{subfig}

\usepackage{mathrsfs}

\newcommand{\bs}[1]{\boldsymbol{#1}}

\usetikzlibrary{arrows.meta, calc, positioning, shadows.blur}

\definecolor{hot1}{RGB}{215,52,56}   
\definecolor{hot2}{RGB}{140,16,16}   
\definecolor{cold1}{RGB}{4,48,159}   
\definecolor{cold2}{RGB}{6,44,112}   
\definecolor{axisgray}{RGB}{40,40,40} 

\tikzset{
  >=Latex,
  axis/.style={line width=1.2pt, draw=axisgray},
  cycle/.style={line width=1.1pt, draw=black},
  dashB/.style={draw=black!60, line width=0.9pt, dash pattern=on 2.2pt off 2.2pt},
  flowarrow/.style={-{Latex[length=4mm]}, line width=.11pt, draw=black},
  heatarrowH/.style={-{Latex[length=4mm]}, line width=2.5pt, draw=hot1},
  heatarrowC/.style={-{Latex[length=4mm]}, line width=2.5pt, draw=cold1},
  pt/.style={circle, inner sep=1.7pt, fill=black},
 resboxH/.style={
    rounded corners=2pt, 
    blur shadow={shadow blur steps=1, shadow xshift=-.1pt, shadow yshift=-0.pt, shadow opacity=0}, 
    inner sep=3.5pt,
    left color=hot2, right color=hot1,shadow scale=0.95, draw=white!70, line width=0.5pt
},
resboxC/.style={
    rounded corners=2pt, 
blur shadow={shadow blur steps=1, shadow xshift=-.1pt, shadow yshift=-0.pt,      shadow scale=0.95, shadow opacity=0.3},
    inner sep=3.5pt,
    left color=cold2, right color=cold1, draw=white!70, line width=0.5pt
}
}

\makeatletter

\begin{document}

\title{Caloric Phenomena and Stirling-Cycle Performance in Heisenberg–Kitaev Magnon Systems}

\author{Bastian Castorene}
\affiliation{Instituto de Física, Pontificia Universidad Católica de Valparaíso, Casilla 4950, 2373223 Valparaíso, Chile}
\affiliation{Departamento de Física, Universidad Técnica Federico Santa María, 2390123 Valparaíso, Chile}

\author{Martin HvE Groves}
\affiliation{Instituto de Física, Pontificia Universidad Católica de Valparaíso, Casilla 4950, 2373223 Valparaíso, Chile}
\affiliation{Departamento de Física, Universidad Técnica Federico Santa María, 2390123 Valparaíso, Chile}

\author{Francisco J. Peña}
\email{francisco.penar@usm.cl}
\affiliation{Departamento de Física, Universidad Técnica Federico Santa María, 2390123 Valparaíso, Chile}

\author{Nicolas Vidal-Silva}
\affiliation{Departamento de Ciencias F\'isicas, Universidad de La Frontera, Casilla 54-D, Temuco, Chile}

\author{Miguel Letelier}
\affiliation{Departamento de Física, Facultad de Ciencias Físicas y Matemáticas, Universidad de Chile, Santiago, Chile}

\author{Roberto E. Troncoso}
\affiliation{Departamento de Física, Facultad de Ciencias, Universidad de Tarapacá, Casilla 7-D, Arica, Chile}

\author{Felipe Barra}
\affiliation{Departamento de Física, Facultad de Ciencias Físicas y Matemáticas, Universidad de Chile, Santiago, Chile}

\author{Patricio Vargas}
\affiliation{Departamento de Física, Universidad Técnica Federico Santa María, 2390123 Valparaíso, Chile}

\begin{abstract}
We investigate the Stirling-cycle performance of a Heisenberg--Kitaev magnonic medium with Dzyaloshinskii--Moriya (DM) interactions. Using linear spin-wave theory, we show the DM interaction preserves spectral symmetry, yielding even caloric responses and symmetric Stirling engine efficiency. In contrast, bond-dependent Kitaev exchange asymmetrically distorts the magnonic density of states, enabling distinct direct and inverse caloric effects. Consequently, Kitaev-driven cycles achieve significantly higher efficiencies than DM-driven protocols, approaching a high-performance saturation regime for negative couplings. This establishes exchange-anisotropic magnets as highly tunable platforms for nanoscale solid-state energy conversion.
\end{abstract}

\maketitle
\section{Introduction}
Magnetic platforms hosting magnons have recently emerged as promising candidates for nanoscale heat management and spin-based energy conversion, owing to their bosonic character, low dissipation, and high degree of tunability \cite{chumak2015magnon,yuan2022quantum,flebus20242024}.
In magnetic insulators, collective spin excitations can be manipulated without
charge transport, enabling thermodynamic cycles that are intrinsically
low-loss and well-suited for caloritronic and quantum-thermodynamic
applications.
A particularly appealing aspect of magnonic systems is that their
thermodynamic response can be controlled not only through conventional
variables such as temperature or magnetic field, but also via microscopic
interaction parameters that shape the excitation spectrum itself
\cite{vidal2024magnonic}.

In this context, magnetic insulators with strong spin--orbit coupling and anisotropic exchange interactions provide a natural platform for
interaction-driven thermodynamics.
Bond-dependent Kitaev exchange and chiral Dzyaloshinskii--Moriya (DM) interactions arise prominently in geometrically frustrated lattices—such as honeycomb, kagomé, and twisted lattices, among others—as well as in iridate-based compounds and layered van der Waals magnets \cite{Jackeli2009,Winter2017,Janssen2019}.
In these materials, the interplay between Heisenberg exchange, anisotropic Kitaev couplings, and DM-induced chirality produces highly tunable
magnonic spectra whose structure can be modified by external fields, strain,
or lattice distortions \cite{Tokura2014,cenker2022reversible,wei2024strain}.

Quantum spin systems as working media for thermal machines has
recently attracted considerable attention \cite{Feldmann2003,Quan2007,Thomas2011,Campisi2016,Kieu2004,Dillenschneider2009,YungerHalpern2019}.
For instance, quantum Otto cycles based on interacting spin systems and
finite clusters have been shown to exhibit strong sensitivity to frustration,
level structure, and magnetic anisotropy.
In particular, Ref.~\cite{Pervez2026} analyzed a quantum Otto heat engine
based on a finite Kitaev--Heisenberg cluster driven by a time-dependent
Zeeman field, demonstrating how the interplay between frustration and
exchange anisotropy can strongly influence the work output and efficiency
of the cycle.
These results highlight the potential of Kitaev-related magnetic systems as
working media for quantum heat engines.

Motivated by this perspective, we investigate the thermodynamic properties of magnons described by a Heisenberg--Kitaev Hamiltonian with second-neighbor Dzyaloshinskii--Moriya (DM) and anisotropic exchange interactions. Since the DM contribution enters the magnon energies quadratically, the energy spectrum and density of states remain invariant under a sign reversal of the coupling. This symmetry dictates that thermodynamic quantities—including entropy, internal energy, specific heat, and caloric response—are even functions of the DM interaction, resulting in strictly symmetric caloric profiles. In contrast, the Kitaev interaction asymmetrically modulates the hopping amplitudes and the magnonic gap, lifting this spectral symmetry and generating distinct direct and inverse caloric effects depending on the sign of the exchange.

Leveraging these results, we design a magnonic Stirling heat engine where either the DM interaction or the Kitaev exchange serves as the control parameter. Cycles driven by the DM coupling inherit the underlying spectral symmetry, yielding efficiency profiles centered at $D=0$. Conversely, Kitaev-driven Stirling cycles exhibit strongly asymmetric efficiency curves and a significant enhancement in work output, reflecting the pronounced redistribution of low-energy spectral weight inherent to bond-dependent anisotropy.

\section{Spin fluctuations}
\begin{figure}[t]
\centering

\includegraphics[width=1\linewidth]{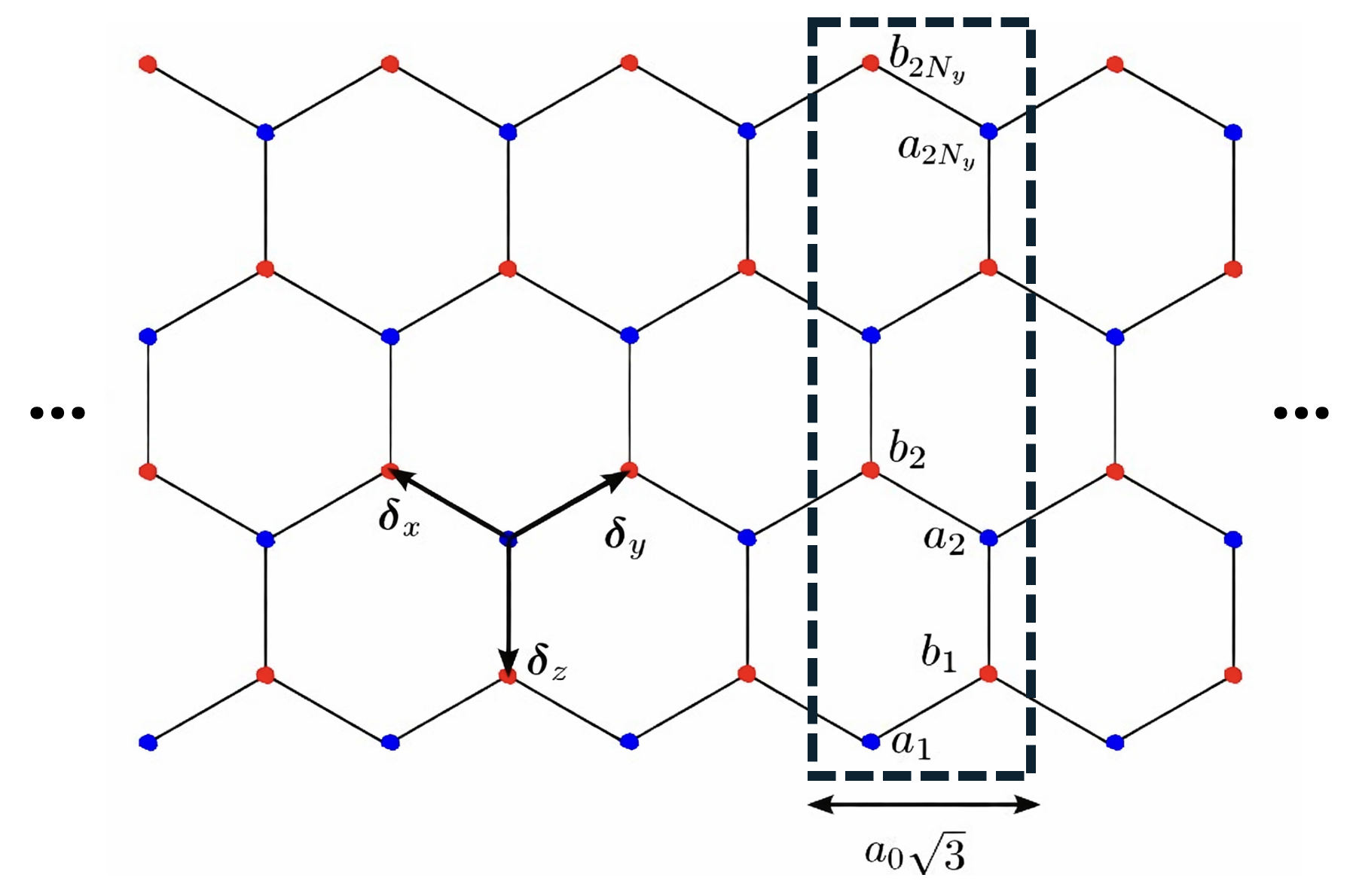}

\caption{Schematic illustration of the magnonic system defined on a hexagonal lattice.
For the numerical calculations, the nanoribbon is periodic along the $x$ direction and finite along the $y$ direction.
The magnetic unit cell, with nearest-neighbor distance $a_0$, is enclosed by a dashed black box,
where the positions of the sites within each sublattice are identified.
The corresponding nearest-neighbor vectors $\boldsymbol{\delta}_x$,
$\boldsymbol{\delta}_y$, and $\boldsymbol{\delta}_z$ are highlighted. The blue and red sites denote the sublattices $\mathcal{A}$ and $\mathcal{B}$, respectively.}
\label{fig:UC}
\end{figure}

We consider a two-dimensional ferromagnetic system of localized spins on a honeycomb lattice, composed of two sublattices, $\mathcal{A}$ and $\mathcal{B}$, represented by the blue and red sites in Fig. \ref{fig:UC}. The spin system is described by the Heisenberg-Kitaev model complemented by an anisotropic Dzyaloshinskii-Moriya interaction. The corresponding spin Hamiltonian is given by
{\color{black}\begin{align}\label{eq:GeneralSpinHamiltonian}
\mathcal{H}[{\bs S}] =& J \sum_{\langle ij\rangle}\nonumber{\boldsymbol{S}}_i\cdot{\boldsymbol{S}}_j-B\sum_{i}S^z_i\\
\nonumber
+&\sum_{\langle\!\langle ij \rangle\!\rangle}
\boldsymbol{D}_{ij}\cdot
\left(
\boldsymbol{S}_i \times \boldsymbol{S}_j
\right)\\ &+2\sum_{\langle ij\rangle_{\gamma}}K^{\gamma}{S}^{\gamma}_i{S}^{\gamma}_j+2\sum_{\langle ij\rangle_{\gamma}}\Gamma^{\gamma}{S}^{\alpha}_i{S}^{\beta}_j .
\end{align}}

The Hamiltonian consists of an isotropic exchange coupling $J<0$ between nearest-neighbor spins at sites $i$ and $j$ on a honeycomb lattice, an external magnetic field $B$ applied along the out-of-plane $z$-axis, and a next-nearest-neighbor Dzyaloshinskii--Moriya interaction (DMI). The DM vector is given by $\boldsymbol{D}_{ij}=D_{ij}\hat{\boldsymbol{z}}$, where $D_{ij}=D\nu_{ij}$ and $\nu_{ij}=\pm 1$ depends on the orientation of the exchange path, taking opposite signs for clockwise and counterclockwise next-nearest-neighbor bonds around each hexagonal plaquette. Consequently, $D$ denotes the strength of the DM interaction. The last two terms correspond to the bond-dependent Kitaev and anisotropic exchange interactions~\cite{joshi2018topological,mcclarty2018topological} with coupling strengths $K^{\gamma}$ and $\Gamma^{\gamma}$, respectively, which naturally arise on the honeycomb lattice. Here, nearest-neighbor bonds are classified into three inequivalent types, denoted by $\gamma \in \{x, y, z\}$ according to their spatial orientation, as illustrated in Fig.~\ref{fig:UC}. On a given $\gamma$ bond, the Kitaev interaction couples the spin components $S_i^\gamma S_j^\gamma$, while the $\Gamma^{\gamma}$ term couples the remaining components $S_i^\alpha S_j^\beta$ with $\{\alpha,\beta,\gamma\}=\{x,y,z\}$.  The Kitaev interaction constitutes a symmetric anisotropic exchange, as it couples equal spin components along bond-dependent directions through terms of the form $S_i^\gamma S_j^\gamma$, which remain invariant under site exchange $i \leftrightarrow j$. In contrast to antisymmetric interactions such as the Dzyaloshinskii--Moriya term, which involves a vector product $\mathbf{S}_i \times \mathbf{S}_j$ and thus changes sign upon exchanging the spin indices, the Kitaev coupling preserves exchange symmetry while introducing strong directional dependence. As a result, it modifies the magnon dispersion by reshaping the spectrum rather than inducing chiral effects.

We introduce quantum spin fluctuations around the out-of-plane ferromagnetic order by performing a local rotation at each spin-lattice site. This magnetic phase can be stabilized by applying high external magnetic fields along the out-of-plane direction, as recently reported \cite{suzuki2021proximate,zhou2023possible,ponomaryov2020nature}. While different magnetic orders can be considered, we here employ linear spin-wave theory (LSWT) to expand the spin operators about the collinear state using the Holstein--Primakoff (HP) transformation \cite{holstein1940field}.
$S^{+}_{\boldsymbol{r}}\approx \sqrt{2S}a_{\boldsymbol{r}}$, $S^{-}_{\boldsymbol{r}} \approx \sqrt{2S}a_{\boldsymbol{r}}^{\dagger}$, and $S^z_{\boldsymbol{r}} = S-a^{\dagger}_{\boldsymbol{r}}a_{\boldsymbol{r}}$, where $a^{\dagger}_{\boldsymbol{r}}$ ($a_{\boldsymbol{r}}$) represents an operator that creates (annihilates) a magnon excitation at position $\boldsymbol{r}$. Next, we replace the HP transformation in the total Hamiltonian, defining the operators $a_{\bs r}$ and $b_{\bs r}$ on the sublattice $\mathcal{A}$ and $\mathcal{B}$, respectively, and keep only the bilinear terms to obtain a tight-binding magnon Hamiltonian ${H}_m$. After Fourier transform, for the field operator $\Psi_{\bs k}=\left(a_{\bs k},b_{\bs k},a^{\dagger}_{-\bs k},b^{\dagger}_{-\bs k}\right)^T$, the magnon Hamiltonian ${\cal H}_m$ is
\begin{align}\label{eq: nondefHamHKG}
{\cal H}_m=\frac{S}{2}\sum_{\bs k}\Psi^{\dagger}_{\bs k}{\cal M}_{\bs k}\Psi_{\bs k}
\end{align}
where the matrix ${\cal M}$ is
\begin{align}
{\cal M}_{\bs k}=\left(\begin{array}{cc}
A_{\bs k}  & B_{\bs k} \\
B^{\dagger}_{\bs k}  & A_{\bs k}
\end{array}\right)
\end{align}
and
\begin{align}
{A}_{\bs k}=\left(\begin{array}{cc}
\kappa_0 + \kappa_{1,\bm k}  & \kappa_{2,-\bs k} \\
\kappa_{2,\bs k}  & \kappa_0-\kappa_{1,\bm k}
\end{array}\right),\quad 
{B}_{\bs k}=\left(\begin{array}{cc}
0  & \kappa_{3,-\bs k} \\
\kappa_{3,\bs k}  & 0
\end{array}\right)   
\label{eq:ABmatrices_0}
\end{align}
where,
\begin{align}
\kappa_0 &= B -3J - 2K^z ,\\
\kappa_{1,\bs k} &=2D\sum_{\bm{\alpha}}\sin\left(\bm{k}\cdot\bm{\alpha}\right),\\
\kappa_{2,\bs k} &= (J+K^x)e^{-i{\bs k}\cdot{\bs \delta_x}}
                  + (J+K^y)e^{-i{\bs k}\cdot{\bs \delta_y}}
                  + J\,e^{-i{\bs k}\cdot{\bs \delta_z}},\\
\kappa_{3,\bs k} &= i\Gamma^z e^{-i{\bs k}\cdot{\bs \delta_z}}
                  + K^x e^{-i{\bs k}\cdot{\bs \delta_x}}
                  - K^y e^{-i{\bs k}\cdot{\bs \delta_y}},
\end{align}
with the link vectors $\bm{\delta}$ given by $\boldsymbol{\delta}_x
= a_0\left(-\frac{\sqrt{3}}{2},\frac{1}{2}\right)$, $\boldsymbol{\delta}_y
= a_0\left(\frac{\sqrt{3}}{2},\frac{1}{2}\right)$, and $\boldsymbol{\delta}_z
= a_0 \,(0,-1)$; and the six next-nearest-neighbors vectors $\bm{\alpha}$ given by $\bm{\alpha}_{1,4} = \pm(\boldsymbol{\delta}_y-\boldsymbol{\delta}_z)
= \pm a_0\left(\frac{\sqrt{3}}{2},\frac{3}{2}\right),$ $\bm{\alpha}_{2,5}=\pm(\boldsymbol{\delta}_x-\boldsymbol{\delta}_z)
= \pm a_0\left(-\frac{\sqrt{3}}{2},\frac{3}{2}\right),$ and $\bm{\alpha}_{3,6}=\pm(\boldsymbol{\delta}_y-\boldsymbol{\delta}_x)
= \pm a_0\left(\sqrt{3},0\right)$, as represented in Fig. \ref{fig:UC}.

For the translationally invariant system, i.e., periodic along $x$ and $y$ directions, the bulk magnon spectrum can be derived by reducing the bosonic Bogoliubov de Gennes Hamiltonian to a $2\times 2$ problem \cite{Shindou2013,Matsumoto2011,AltlandSimons}. In the following, we restrict to the case $\Gamma^z = 0$, which is also the regimen used in the numerical implementation (see the Appendix.~\ref{Bulk_magnon} for further details). Thus, the magnon branches are given by $\varepsilon_{\bm{k},\pm}=\sqrt{\epsilon_{\bm{k},\pm}^{2}}$, where:
\begin{align}
\epsilon_{\bm{k},\pm}^{2}=\kappa_0^2+\kappa_{1,\bm{k}}^2+\vert\kappa_{2,\bm{k}}\vert^2-\vert\kappa_{3,\bm{k}}\vert^2\pm\sqrt{\Delta_{\bm{k}}},
\label{eq:magnondispersion}
\end{align}
with
\begin{align} \nonumber\Delta_{\bm{k}}&=4\kappa_0^2\kappa_{1,\bm{k}}^2+4\kappa_0^2\vert \kappa_{2,\bm{k}}\vert^2-4\kappa_{1,\bm{k}}^2\vert\kappa_{3,\bm{k}}\vert^2\\
    &+\left(\kappa_{2,\bm{k}}\kappa_{3,-\bm{k}}-\kappa_{2,-\bm{k}}\kappa_{3,\bm{k}}\right)^2.
    \label{eq:determ}
\end{align}
One can readily notice two main aspects from the above expressions. First, The magnon energy spectrum is invariant under a sign reversal of the DM interaction. Note that, this interaction enters only in $\kappa_{1,\bm{k}}$, which is linear in the $D$ parameter. However, $\kappa_{1,\bm{k}}$ contributes to the magnon energies only through quadratic combinations. As a result, $\varepsilon_{\bm{k}}^{\pm}(D) =\varepsilon_{\bm{k}}^{\pm}(-D)$ and one should expect symmetric caloric response in these systems \cite{vidal2024magnonic}.
Contrarily, if we take for simplicity $K^x=K^y=K^z=K$, one can notice that $\varepsilon_{\bm{k}}^{\pm}(K) \neq \varepsilon_{\bm{k}}^{\pm}(-K)$. Indeed, the asymmetry originates from the terms $|\kappa_{2,\bm{k}}|^2$ and $\kappa_0^2$, where the Kitaev coupling $K$ enters linearly.   Consequently, we expect distinct caloric responses when varying the Kitaev and DM interactions across positive and negative values. \\

We employ a ribbon geometry with a finite width of $\mathcal{N}_s$ sites along the transverse direction. To avoid the influence of edge states, the width is chosen to be sufficiently large to reproduce the bulk dispersion, providing a consistent sampling of the bulk spectrum. Thus, for a semi-infinite system (periodic along a given direction and finite along the other one, see Fig. \ref{fig:UC}), the magnon Hamiltonian (Eq. \ref{eq: nondefHamHKG}) can be numerically (para)diagonalized by employing Colpa's algorithm \cite{colpa1978diagonalization}. We can write down
\begin{align}
    \mathcal{H}_m=\frac{S}{2}\sum_{\bm{k}}[\alpha_{\bm k}^{\dagger}\hspace{0.1cm}\alpha_{-\bm{k}}]\mathcal{M}_{\bm k}[\alpha_{\bm k}\hspace{0.1cm}\alpha_{-\bm{k}}^{\dagger}],
\end{align}
with $\alpha_{\bm k}\equiv (a_{\bm{k}1},a_{\bm{k}2},\ldots,a_{\bm{k}N_y},\hspace{0.1cm}b_{\bm{k}1},b_{\bm{k}2},\ldots,b_{\bm{k}N_y})$, where $N_y$ denotes the number of sites in each sublattice of the unit cell (see Fig.~\ref{fig:UC}). Colpa's algorithm will return us a paraunitary matrix $\mathcal{T}_{\bm k}$ that satisfies
\begin{align}
   \mathcal{T}_{\bm k}^{\dagger}\mathcal{M}_{\bm k}\mathcal{T}_{\bm k}=
   \begin{pmatrix} 
   \mathcal{E}_{\bm k}& 0\\
   0 &\mathcal{E}_{-\bm k}
   \end{pmatrix}
\end{align}
where $\mathcal{E}_{\bm k}$ is a $2N_y\times 2N_y$ diagonal matrix containing the eigenenergies $\varepsilon_{\bm{k},n}$ for the \textit{n}-th-magnon branch. The respective eigenvectors are written as
\begin{align}
     \begin{pmatrix} 
   \alpha_{\bm k}\\
   \alpha_{-\bm k}^{\dagger}
   \end{pmatrix}=
    \mathcal{T}_{\bm k}\begin{pmatrix} 
   \gamma_{\bm k}\\
   \gamma_{-\bm k}^{\dagger}
   \end{pmatrix},
\end{align}
being $\gamma_{\bm k}$ ($\gamma_{\bm k}^{\dagger}$) the magnon annihilation (creation)
operator written in the diagonal basis. The paraunitary matrix $\mathcal{T}_{k}$ is decomposed as
\begin{align}
   \mathcal{T}_{k} = 
   \begin{pmatrix}
       U_{\bm k}& V_{-\bm k}^{*} \\
       V_{\bm k}& U_{-\bm k}^{*}
   \end{pmatrix},
\end{align}
where matrices $U_{\bm k}$ and $V_{\bm k}$ corresponds to the coefficient of the generalized Bogoliubov transformation \cite{colpa1978diagonalization}. The magnon wave function belonging to the \textit{n}-th band is defined as $\psi_{\bm{k},n}=\alpha_{\bm{k}n}^{\dagger}|GS\rangle$, being $|GS\rangle$ the magnon ground state. Thus, in the diagonal eigenbasis $\{\alpha^{\dagger}_{\bs k},\alpha_{\bs k}\}$, the Hamiltonian reads ${\cal H}_m=\sum_{\bs k, n}\varepsilon_{\bs k, n}\alpha^{\dagger}_{\bs k n}\alpha_{\bs k n}$, with the $\varepsilon_{\bs k, n}$ the eigen-energies for the \textit{n}-th magnon mode. For the numerical implementation, we set $a_0=1$, $J=-1$, $B=1$, $\Gamma=0$, and $S=1$, and vary the relevant parameters $D$ and $K_x=K_y=K_z\equiv K$ within the ranges $-0.4 < D < 0.4$ and $-0.4 < K < 0.4$, respectively. Throughout this work, energies are expressed in units of $|J|$.


\section{Bands and Density of States}

\begin{figure}[t]
\centering

\includegraphics[width=0.90\linewidth]{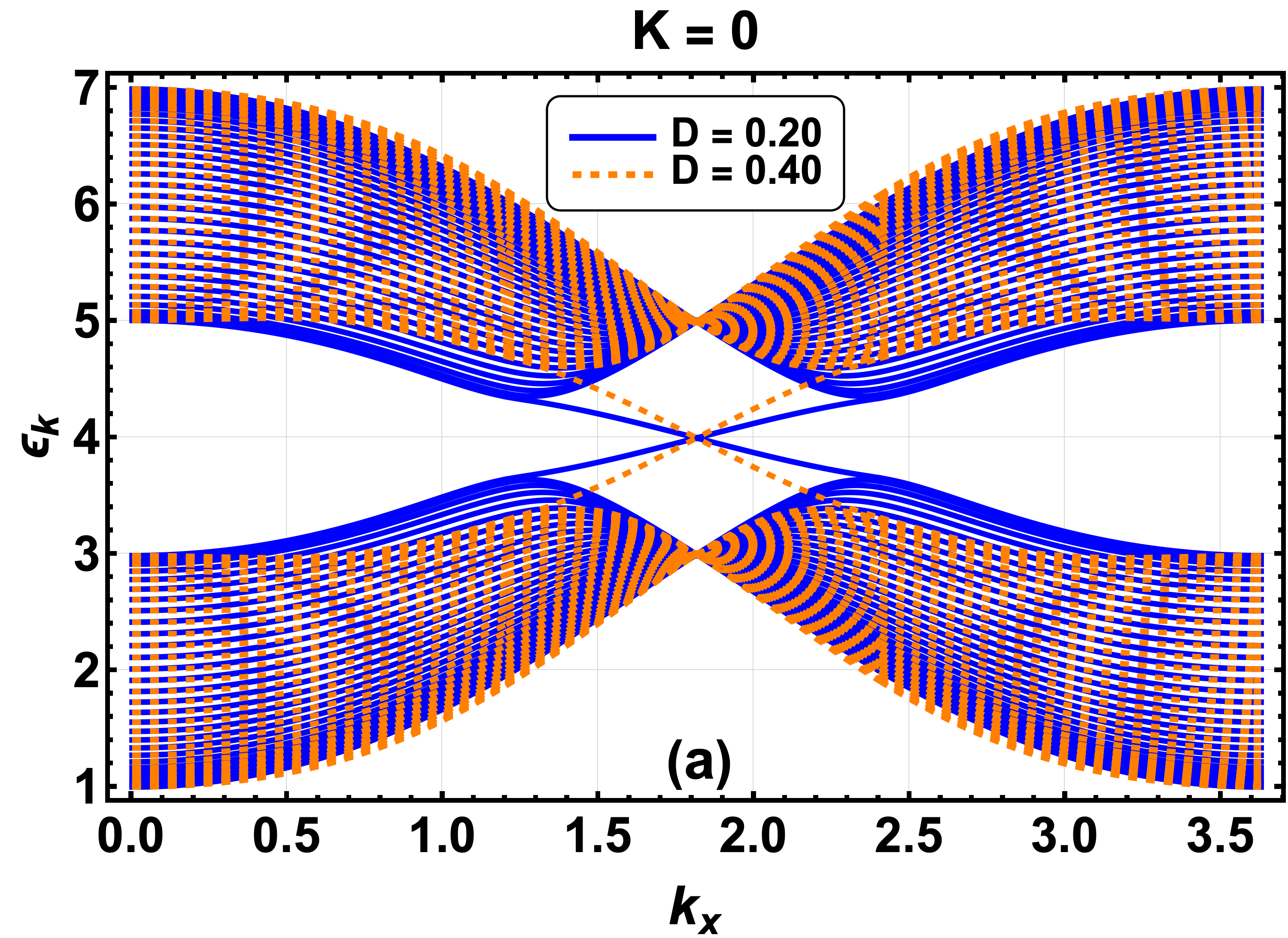}

\vspace{0.8em}

\includegraphics[width=0.90\linewidth]{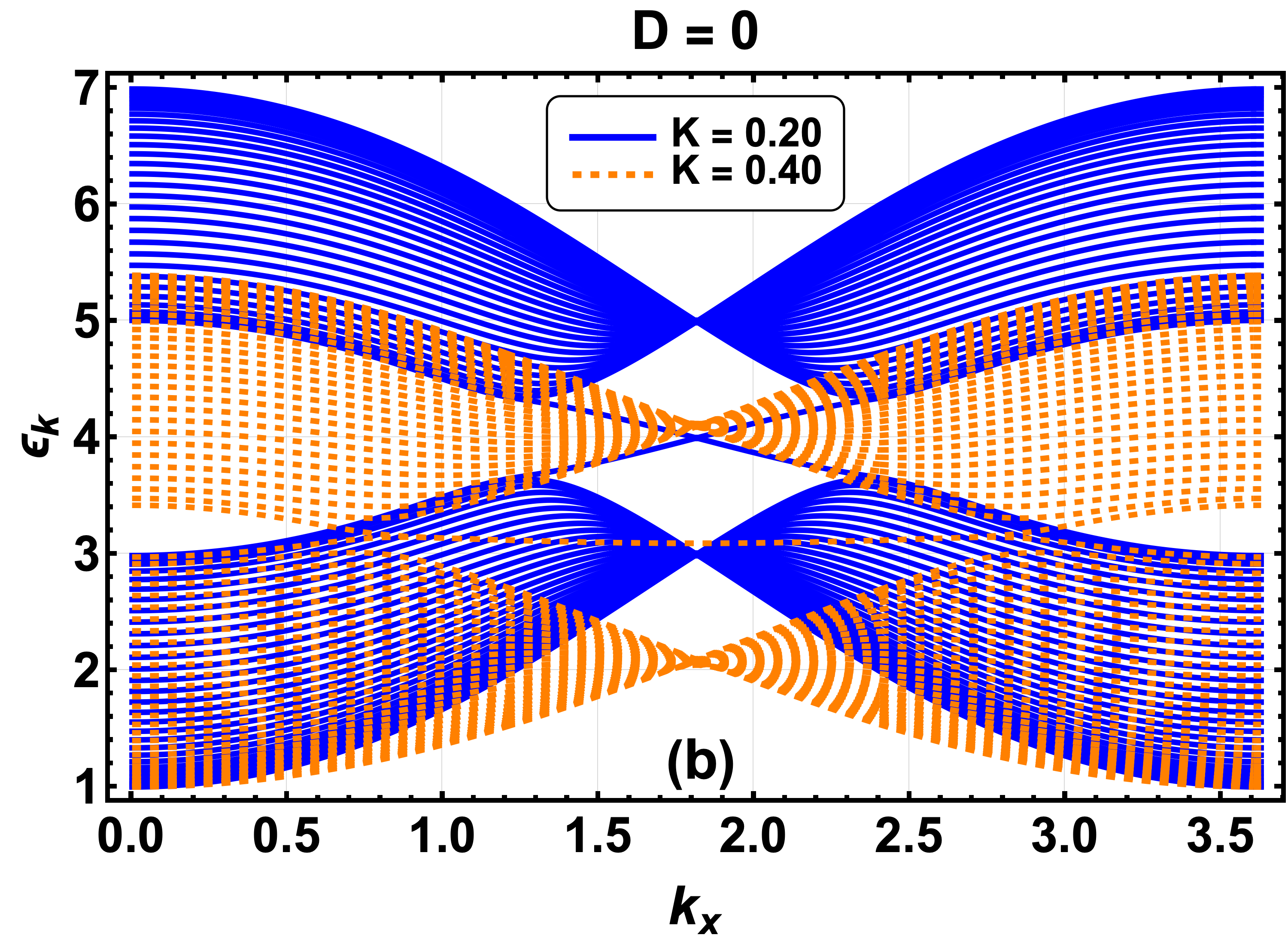}

\caption{(Color online) Magnon band structure $\varepsilon_{\bm k,n}$ for $\mathcal{N}_s=30$.
(a) Fixed $K=0$ and two values of the DM interaction: $D=0.20$ (solid) and $D=0.40$ (dashed).
(b) Fixed $D=0$ and two values of the Kitaev exchange: $K=0.20$ (solid) and $K=0.40$ (dashed).}
\label{fig:bands_stacked_final}
\end{figure}

Fig.~\ref{fig:bands_stacked_final} illustrates representative magnon dispersions $\varepsilon_{\bs k, n}$ obtained via paraunitary diagonalization of the linear spin-wave Hamiltonian for $\mathcal{N}_s=30$ sites. Specifically, Fig.~\ref{fig:bands_stacked_final}(a) presents the $K=0$ case for two values of the second-neighbor DM coupling, $D=0.20$ and $D=0.40$, while Fig.~\ref{fig:bands_stacked_final}(b) fixes $D=0$ to compare $K=0.20$ and $K=0.40$. Qualitatively, varying $D$ at $K=0$ preserves the overall shape of the dispersions and primarily shifts spectral features smoothly. In contrast, varying $K$ at $D=0$ induces a more pronounced reshaping of the band structure. Given that the thermodynamic response of a bosonic working medium is governed by the available phase space of low-energy excitations, these spectral differences suggest that Kitaev-driven protocols can generate stronger thermodynamic signatures than DM-driven ones, a result quantified in the subsequent sections.

The magnon spectrum comprises two branches, yielding $N_{\mathrm{states}} = 2\mathcal{N}_s = 60$ single-particle states. We normalize the density of states (DOS) $g(\varepsilon;D,K)$ such that its integral over the spectral range $[\varepsilon_{\min}, \varepsilon_{\max}]$ reproduces this total:
\begin{equation}
\int_{\varepsilon_{\min}}^{\varepsilon_{\max}} g(\varepsilon;D,K)\, d\varepsilon = N_{\mathrm{states}}.
\end{equation}
This normalization ensures thermodynamic consistency across all simulated configurations.

\begin{figure}[t]
\centering
{%
\includegraphics[width=1.0\linewidth]{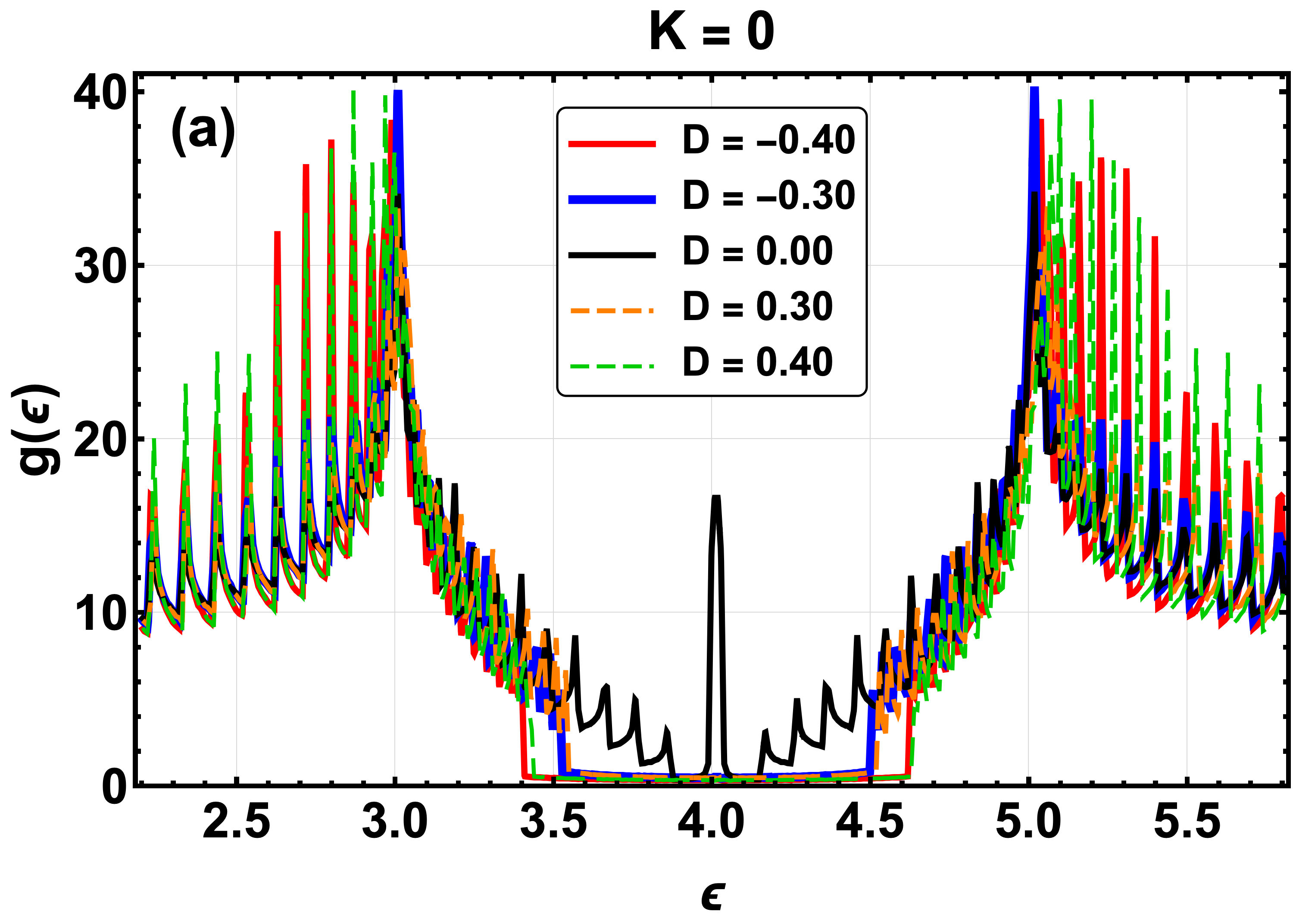}%
}\\
\vspace{.5cm}
{%
\includegraphics[width=1.0\linewidth]{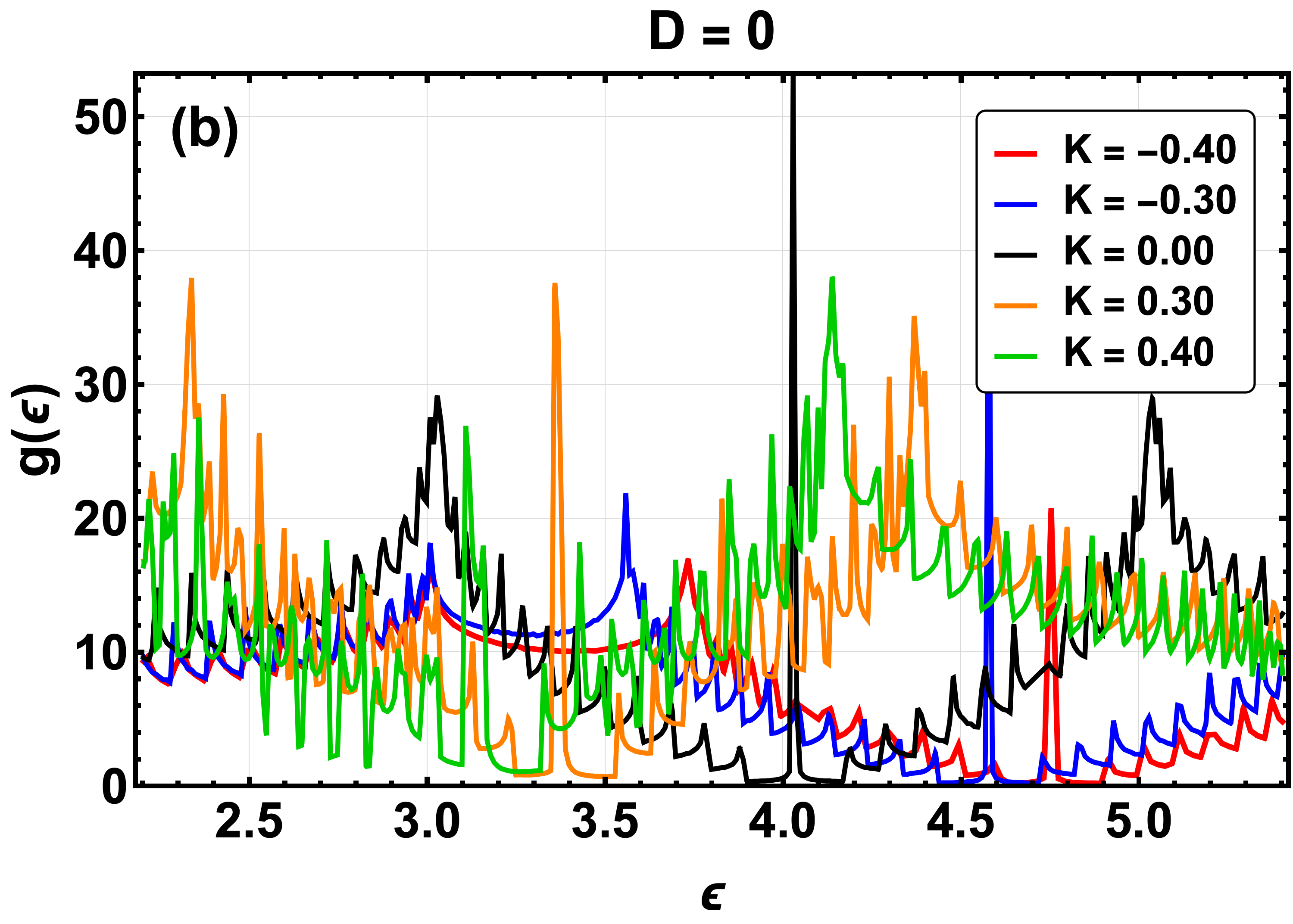}%
}

\caption{(Color online) Normalized magnon density of states $g(\varepsilon)$ used in the thermodynamic calculations:
(a) $g(\varepsilon)$ at fixed $K=0.00$ for $D=\{0.00,\pm 0.30,\pm 0.40\}$ (positive $D$ shown with solid lines and
negative $D$ with dotted lines); (b) $g(\varepsilon)$ at fixed $D=0.00$ for $K=\{0.00,\pm 0.30,\pm 0.40\}$
(positive $K$ shown with solid lines and negative $K$ with dotted lines), truncated at $\varepsilon \le 5.2$ for
numerical stability near the upper band edge. In both panels the DOS is normalized to the total number of
single--particle states $N_{\mathrm{states}}=2\mathcal{N}_s=60$}
\label{fig:DOS_panels}
\end{figure}

Fig~\ref{fig:DOS_panels} illustrates the energy-resolved counterparts to the dispersion trends. In Fig.~\ref{fig:DOS_panels}(a) At $K=0$, the DOS remains invariant under $D \to -D$ and shows only moderate redistributions, as the DM coupling primarily modifies hopping phases without fundamental spectral reshaping. Conversely in Fig.~\ref{fig:DOS_panels}(b), at $D=0$, the Kitaev exchange induces a pronounced redistribution of spectral weight, particularly in the low-energy sector. Notably, the DOS is asymmetric under $K \to -K$, yielding distinct spectral profiles for the same $|K|$. Since bosonic observables are sensitive to low-energy excitations, these DOS asymmetries directly drive the enhanced and direction-dependent caloric responses and engine performance characterized in the following sections.


\section{Thermodynamic quantities}
\label{sec:thermo}

Once the bosonic excitation spectrum
$\{\varepsilon_n(\bm{k};D,K)\}$
is obtained from the spin Hamiltonian, the thermodynamic properties of the magnon gas can be evaluated within the grand-canonical formalism. We consider noninteracting bosonic quasiparticles whose dispersion depends parametrically on $(D,K)$ couplings. 

To regularize the Bose occupation near the band edge, we set the chemical potential slightly below the minimum energy of the lowest magnon band,
\begin{equation}
\mu=\varepsilon_{\min}-\delta,
\qquad \delta>0,\qquad \delta\ll 1,
\end{equation}
where $\varepsilon_{\min}$ denotes the bottom of the magnon spectrum for the chosen set of parameters $(D,K)$. In practice, this means that $\mu$ is always taken infinitesimally below the band minimum. This prescription prevents Bose-Einstein condensation within the temperature range considered and ensures that the thermal population is dominated by low-energy magnon excitations close to the band edge.

It is important to note that, in magnetic systems, the number of magnons is generally not strictly conserved due to interactions with the lattice and other relaxation processes. Consequently, the chemical potential introduced here should be interpreted as an effective parameter regulating the bosonic occupation of the magnon spectrum within the grand-canonical formalism, rather than as a conserved thermodynamic chemical potential \cite{demokritov2006bose,cornelissen2016magnon}. This procedure is commonly adopted in thermodynamic descriptions of bosonic quasiparticles, such as magnons or photons, where the chemical potential primarily serves as a reference energy controlling the thermal population of the excitation spectrum .

Although the band minimum $\varepsilon_{\min}$ generally depends on the microscopic parameters $D$, $K$, all thermodynamic derivatives defining response functions and generalized forces are evaluated at fixed $\mu$, in accordance with the natural variables of the grand potential $\Omega(T,\mu;D,K)$. Operationally, $\mu$ acts as a fixed reference energy during the evaluation of partial derivatives such as $(\partial\Omega/\partial D)_{T,\mu,K}$ and $(\partial\Omega/\partial K)_{T,\mu,D}$. This prescription guarantees the internal consistency of the thermodynamic framework and of the generalized first law, while keeping the chemical potential sufficiently close to the band edge to capture the spectral region that dominates the entropy and caloric response.
Within this framework, the grand potential is given by
\begin{equation}
\Omega = k_B T \int d\epsilon\, g(\epsilon; D, K) \ln\!\left[1 - e^{-\beta(\epsilon-\mu)}\right],
\label{eq:grand_potential}
\end{equation}
where $\beta = 1/(k_B T)$ and $\mu$ is the chemical potential. The internal energy $U$ and entropy $S$ are expressed as
\begin{equation}
U = \int d\epsilon\, g(\epsilon; D, K) \epsilon\, n_B ,
\label{eq:internal_energy}
\end{equation}
\begin{equation}
S = k_B \int d\epsilon\, g(\epsilon; D, K) \Big[(1 + n_B)\ln(1 + n_B) - n_B \ln n_B\Big],
\label{eq:entropy}
\end{equation}
where $n_B  = [e^{\beta(\epsilon-\mu)} - 1]^{-1}$ is the Bose-Einstein distribution. The specific heat at constant $\mu$ follows the fluctuation-dissipation form
\begin{equation}
C = \int d\epsilon\, g(\epsilon; D, K) \frac{(\epsilon-\mu)^2}{k_B T^2} n_B(1+n_B).
\label{eq:specific_heat}
\end{equation}

Treating $D$ and $K$ as external control parameters, we define their conjugate generalized forces as
\begin{equation}
X_D = -\left( \frac{\partial\Omega}{\partial D} \right)_{T,\mu,K }, \quad X_K = -\left( \frac{\partial\Omega}{\partial K} \right)_{T,\mu,D },
\label{eq:forces}
\end{equation}
which quantify the work associated with variations in the microscopic couplings. The thermodynamic relations are summarized in the generalized first law:
\begin{equation}
dU = T dS + \mu dN + X_D dD + X_K dK,
\label{eq:firstlaw}
\end{equation}
where $N$ is the total magnon number. This framework enables the analysis of heat cycles driven by the modulation of both temperature and exchange interactions.


\section{Stirling cycle driven by a coupling parameter}
\label{subsec:stirling_lambda}

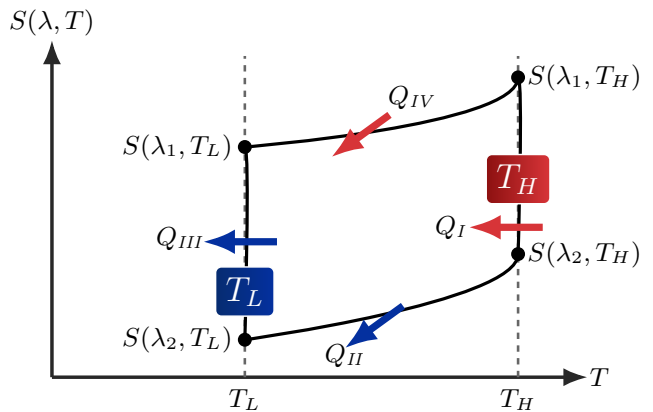
\begin{figure}
\centering
\resizebox{1.\linewidth}{!}{\begin{tikzpicture}[x=0.65cm,y=0.65cm]

\draw[axis,->] (1,0) -- (11,0) node[right=-3pt] {$T$};
\draw[axis,->] (1,0) -- (1,6.3) node[above=-2pt] {$S(\lambda,T)$};

\def\TL{4.6}
\def\TH{9.7}

\draw[dashB] (\TL,0) -- (\TL,6);
\draw[dashB] (\TH,0) -- (\TH,6);

\coordinate (A) at (\TH,5.6);
\coordinate (B) at (\TH,2.3);
\coordinate (C) at (\TL,.7);
\coordinate (D) at (\TL,4.3);

\draw[cycle] (A) .. controls ($(A)+(0.1,0)$) and ($(B)$) .. (B);
\draw[cycle] (B) .. controls ($(B)+(0,-.9)$) and ($(C)$) .. (C);
\draw[cycle] (C) .. controls ($(C)$) and ($(D)+(0.1,0)$) .. (D);
\draw[cycle] (D) .. controls ($(D)$) and ($(A)+(0,-.9)$) .. (A);

\node[right=0pt of A] {$S(\lambda_1,T_H)$};
\node[right=0pt of B] {$S(\lambda_2,T_H)$};
\node[left=1pt of C] {$S(\lambda_2,T_L)$};
\node[left=1pt of D] {$S(\lambda_1,T_L)$};

\node[below] at (\TL,-0.05) {$T_L$};
\node[below] at (\TH,-0.05) {$T_H$};

\node[resboxH] (TH) at (9.7,3.7) {\large $\textcolor{white}{T_H}$};
\node[resboxC] (TL) at (4.6,1.6) {\large $\textcolor{white}{T_L}$};

\draw[heatarrowH] ($(A)!-.!(B)$) ++(0.46,-2.8) -- ++(-1.4,0)
node[pos=1.19, above left=-12pt] {$Q_{\textit{I}}$};

\draw[heatarrowC] ($(B)!0.48!(C)$) ++(.3,-.2) -- ++(-1.1,-0.8)
node[pos=1, below=-6pt] {$Q_{\textit{II}}$};

\draw[heatarrowC] ($(C)!0.55!(D)$) ++(0.6,-0.15) -- ++(-1.4,0)
node[pos=1.1, above left=-10pt] {$Q_{\textit{III}}$};

\draw[heatarrowH] (7.3,4.9) -- ++(-1.1,-.8)
node[pos=-1, below left=1pt] {$Q_{\textit{IV}}$};

\foreach \P in {A,B,C,D} \fill (\P) node[pt] {};

\end{tikzpicture}}
\caption{Stirling cycle in the $S(\lambda,T)$--$T$ plane for a bosonic working medium, where the control parameter $\lambda\in\{K,D\}$ is modulated quasistatically. The vertical branches correspond to the isothermal strokes at $T_H$ and $T_L$, whereas the upper and lower curved branches correspond to the isoparametric processes at fixed $\lambda_1$ and $\lambda_2$, respectively. Red arrows denote heat absorbed by the working medium, whereas blue arrows indicate heat released to the reservoirs.}
\label{fig:stirling_engine}
\end{figure}

We now consider a Stirling cycle described in Fig.~\ref{fig:stirling_engine}, where work is produced by quasistatic modulation of a single coupling $\lambda\in\{K,D\}$, while the remaining parameters are kept fixed.
The cycle consists of two isothermal strokes at temperatures $T_H>T_L$ and two isoparametric strokes at fixed $\lambda$.

\paragraph*{Cycle definition.}
Let $\lambda_1<\lambda_2$ denote the extrema of the control parameter.

\paragraph*{\textbf{(I) Isothermal expansion at $T_H$.}}
The control parameter is varied as $\lambda:\lambda_1\to\lambda_2$ at fixed temperature $T=T_H$. The absorbed heat and the generalized work are
\begin{equation}
Q_I=T_H\,\Delta S_{\rm iso}(T_H;\lambda_1\to\lambda_2),
\quad
W_I=\int_{\lambda_1}^{\lambda_2}X_\lambda(T_H,\lambda)\,d\lambda .
\label{eq:stroke_I}
\end{equation}

\paragraph*{\textbf{(II) Isoparametric cooling at $\lambda_2$.}}
The temperature is reduced from $T_H$ to $T_L$ at fixed $\lambda=\lambda_2$. Since $d\lambda=0$, no generalized work is performed, and thus
\begin{equation}
W_{II}=0,
\qquad
Q_{II}=\int_{T_H}^{T_L} C(T;\lambda_2)\,dT .
\label{eq:stroke_II}
\end{equation}

\paragraph*{\textbf{(III) Isothermal compression at $T_L$.}}
The control parameter is swept back as $\lambda:\lambda_2\to\lambda_1$ at fixed temperature $T=T_L$. The released heat and the associated work are
\begin{equation}
Q_{III}=T_L\,\Delta S_{\rm iso}(T_L;\lambda_2\to\lambda_1),
\quad 
W_{III}=\int_{\lambda_2}^{\lambda_1}X_\lambda(T_L,\lambda)\,d\lambda .
\label{eq:stroke_III}
\end{equation}

\paragraph*{\textbf{(IV) Isoparametric heating at $\lambda_1$.}}
The temperature increases from $T_L$ to $T_H$ at fixed $\lambda=\lambda_1$. Again, no generalized work is performed, so that
\begin{equation}
W_{IV}=0,
\qquad
Q_{IV}=\int_{T_L}^{T_H} C(T;\lambda_1)\,dT .
\label{eq:stroke_IV}
\end{equation} 
The net work $W = \sum W_i$ and the total absorbed heat $Q_{\text{in}} = Q_I + Q_{IV}$ determine the thermal efficiency:
\begin{equation}
\eta=\frac{W}{Q_{\rm in}}
 .
\label{eq:efficiency}
\end{equation}
Throughout this work we do not assume the presence of an ideal regenerator.
Accordingly, the heat exchanged along the isoparametric strokes,
$Q_{II}$ and $Q_{IV}$, is treated as ordinary heat transfer with the thermal
reservoirs and is fully included in the thermodynamic balance.

In the present magnonic working medium, the control parameter $\lambda\in\{K,D\}$ reshapes the low-energy sector of the spin-wave spectrum and, consequently, the density of states.
At low temperatures, the entropy is dominated by the population of low-energy
magnon modes, so parameter variations that enhance the spectral weight near the
band bottom or soften the lowest excitations produce a rapid increase of
$S(\lambda,T)$ and a larger isothermal entropy change.
Within the Heisenberg--Kitaev--Dzyaloshinskii--Moriya model studied here, this
mechanism is particularly effective under modulation of the Kitaev exchange.
As shown in Figs.~\ref{fig:bands_stacked_final} and~\ref{fig:DOS_panels},
changes in $K$ induce a pronounced reshaping of the magnon dispersion and a
strong redistribution of the density of states toward low energies. By
contrast, variations of the DM coupling primarily generate smoother and largely
symmetric spectral modifications, resulting in a comparatively weaker caloric
response.

\begin{figure}[t]
    \centering
    \includegraphics[width=1.0\linewidth]{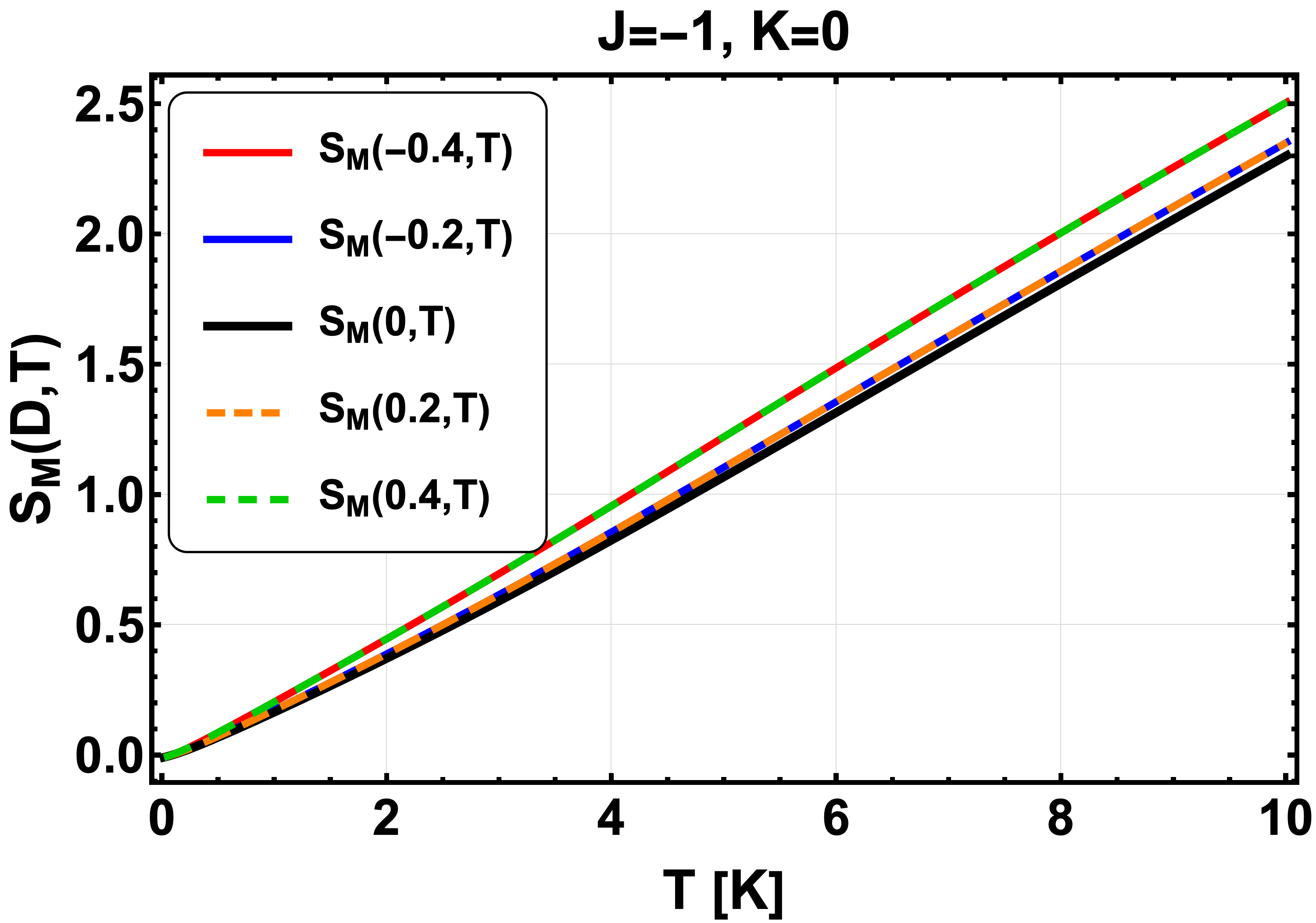}
    \caption{Magnonic entropy \( S_M(D,T) \) as a function of temperature for different values of the Dzyaloshinskii–Moriya interaction \( D \), with \( J = -1 \) and \( K = 0 \). }
    \label{fig:entropy_vs_T}
\end{figure}
\begin{figure}[t]
    \centering
    \includegraphics[width=0.95\linewidth]{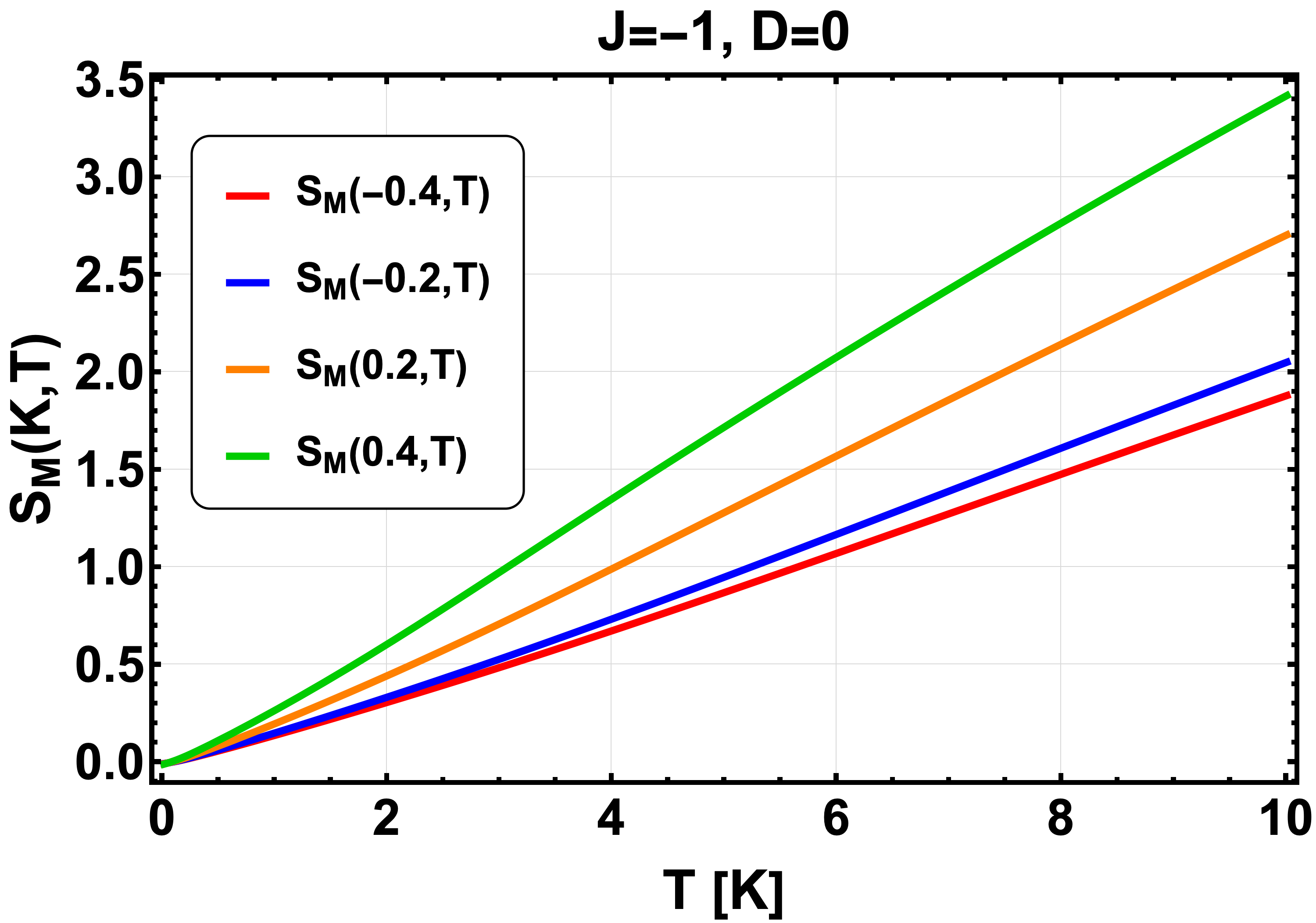}
    \caption{Magnonic entropy \(S_M(T)\) for different values of the Kitaev coupling \(K\) at fixed \(J=-1\) and \(D=0\). In contrast to the DM-induced case, the entropy shows a clear asymmetry around \(K=0\), reflecting the direct modification of the magnon spectrum by the bond-dependent Kitaev interaction.}
    \label{fig:entropy_vs_K}
\end{figure}

From an engine perspective, increasing the control parameter $\lambda$ during the isothermal expansion favors spectral configurations with enhanced low-energy weight. Such configurations maximize the entropy at a given temperature, thereby enlarging the area enclosed by the Stirling cycle in the $X_\lambda$--$\lambda$ plane and increasing the work output. Here, $\lambda$ represents either the Kitaev exchange ($\lambda=K$) or the DM interaction ($\lambda=D$), with the complementary coupling held constant. The analysis of $X_\lambda$--$\lambda$ loops and the isothermal entropy change $\Delta S_{\text{iso}}(T)$ provides a comprehensive characterization of the engine's performance. 

Finally, we clarify that the entropy $S$ defined in Eq.~(\ref{eq:entropy}) accounts exclusively for the contribution of the bosonic magnon bands. Therefore, all thermodynamic results discussed hereafter refer only to this magnonic component, which we denote as $S_M$.

\section{Thermodynamics Results}

In this section, we analyze the thermodynamic properties of the magnonic system, beginning with the entropy $S_{M}$, which determines the heat exchanged during the isothermal strokes of the Stirling cycle. we will examine adiabatic trajectories as a diagnostic tool to reveal symmetry properties of the model and to characterize how the different microscopic couplings reshape the underlying magnonic spectrum. This combined analysis provides a clear understanding of how entropy encode the macroscopic thermodynamic response arising from the microscopic interactions.

Fig.~\ref{fig:entropy_vs_T} illustrates the temperature dependence of the magnonic entropy $S_M(T)$ for several values of $D$ at fixed $K=0$. The entropy grows monotonically with temperature and, notably, exhibits an exact symmetry under the transformation $D \to -D$. This even dependence arises because the DM interaction contributes only a complex phase to the magnon hopping amplitudes, leaving the energy spectrum and the density of states invariant. Consequently, all thermodynamic quantities derived from the spectrum remain unchanged under a sign reversal of $D$.

In contrast, Fig.~\ref{fig:entropy_vs_K} shows the entropy as a function of temperature for various values of the Kitaev coupling $K$ at $D=0$. Unlike the DM case, these curves exhibit a pronounced asymmetry under $K \to -K$. This reflects the fact that the bond-dependent Kitaev interaction modifies the real hopping amplitudes in a non-trivial manner, thereby reshaping the magnonic spectrum and the density of states. Ultimately, while the DM interaction—antisymmetric under spin exchange—leads to a symmetric entropy response, the anisotropic Kitaev exchange, which is symmetric under spin exchange, produces an asymmetric thermodynamic fingerprint due to its capacity to distort the magnonic spectral weight.

\begin{figure}[t]
    \centering
    \includegraphics[width=0.95\linewidth]{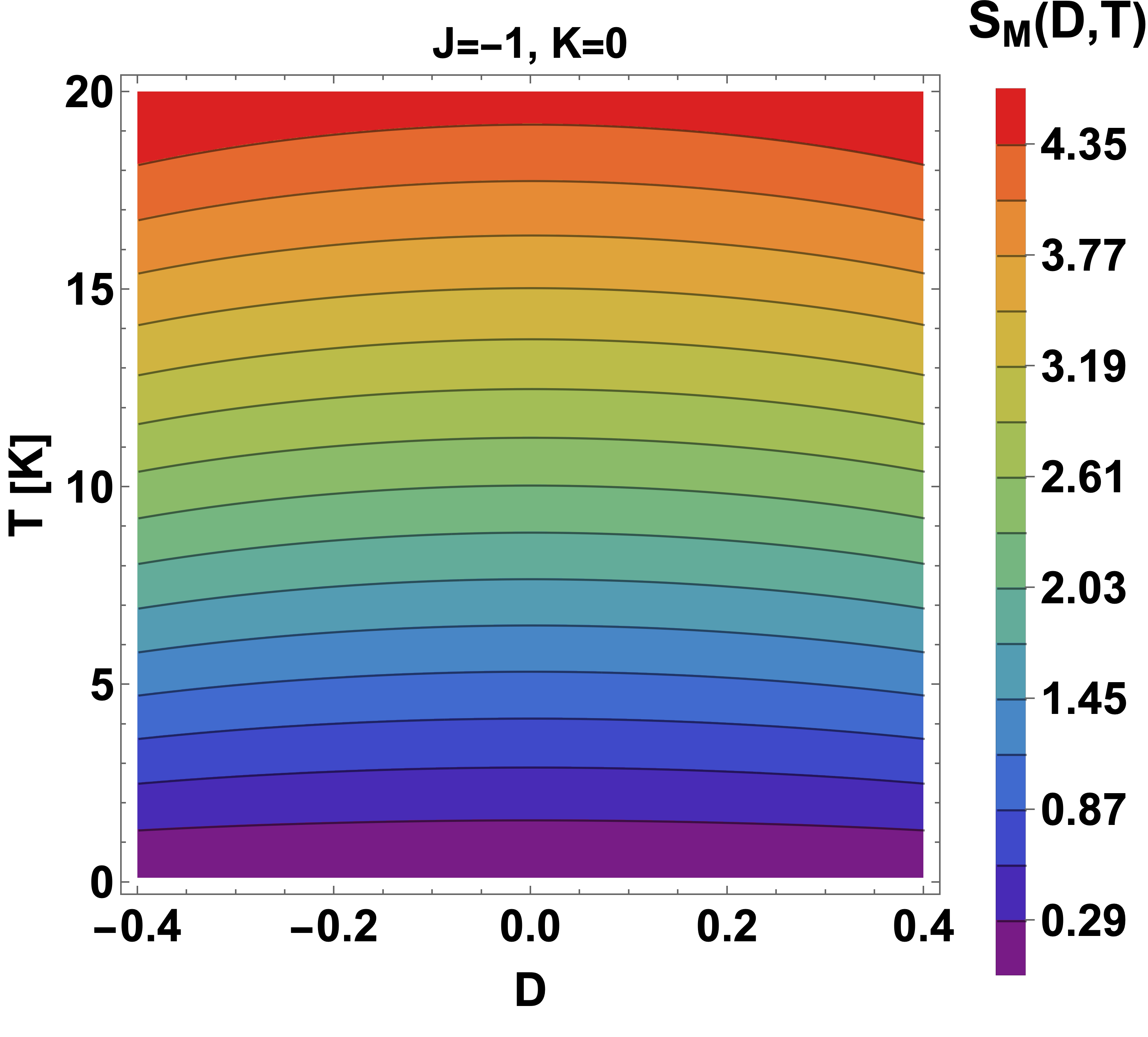}
    \caption{Constant–entropy contour map \(S_M(D,T)\) for several values of the Dzyaloshinskii--Moriya coupling \(D\), with \(J=-1\) and \(K=0\). The contours show perfect symmetry under \(D \to -D\), reflecting the invariance of the magnonic spectrum with respect to the sign of the second-neighbor DM interaction.}
    \label{fig:entropy_contour_D}
\end{figure}
Fig.~\ref{fig:entropy_contour_D} provides a two-dimensional perspective of the same symmetry through the contour map \(S_M(D,T)\). All isoentropy lines exhibit mirror symmetry around \(D=0\), confirming that the underlying magnonic spectrum is strictly even in the DM coupling.
\begin{figure}[t]
    \centering
    \includegraphics[width=0.95\linewidth]{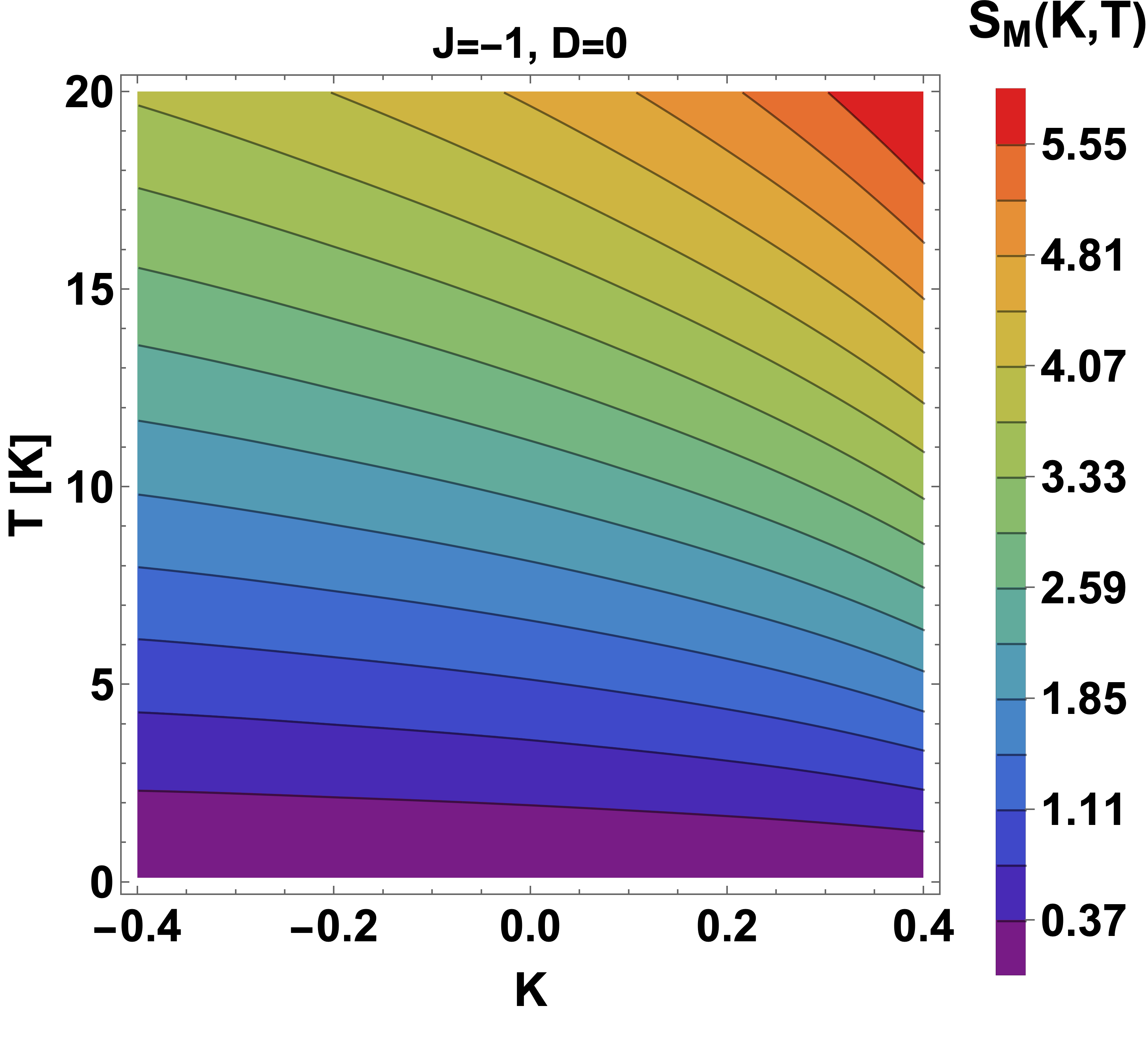}
    \caption{Constant–entropy contour map \(S_M(K,T)\) for several values of the Kitaev coupling \(K\), with \(J=-1\) and \(D=0\). In contrast to the DM case, the contours exhibit a strong asymmetry under \(K \to -K\), highlighting the non-symmetric modification of the magnonic spectrum induced by the bond–dependent Kitaev exchange.}
    \label{fig:entropy_contour_K}
\end{figure}
Finally, the contour representation in Fig.~\ref{fig:entropy_contour_K} highlights the directional dependence of the isoentropy lines when varying the Kitaev coupling. The absence of any symmetry under \(K\to -K\) directly reflects the nontrivial reshaping of the dispersion relation produced by the bond–anisotropic Kitaev exchange, in sharp contrast with the phase-only effect of the DM interaction.

\subsection{Caloric Response}

\begin{figure}[t]
    \centering
    \includegraphics[width=0.95\linewidth]{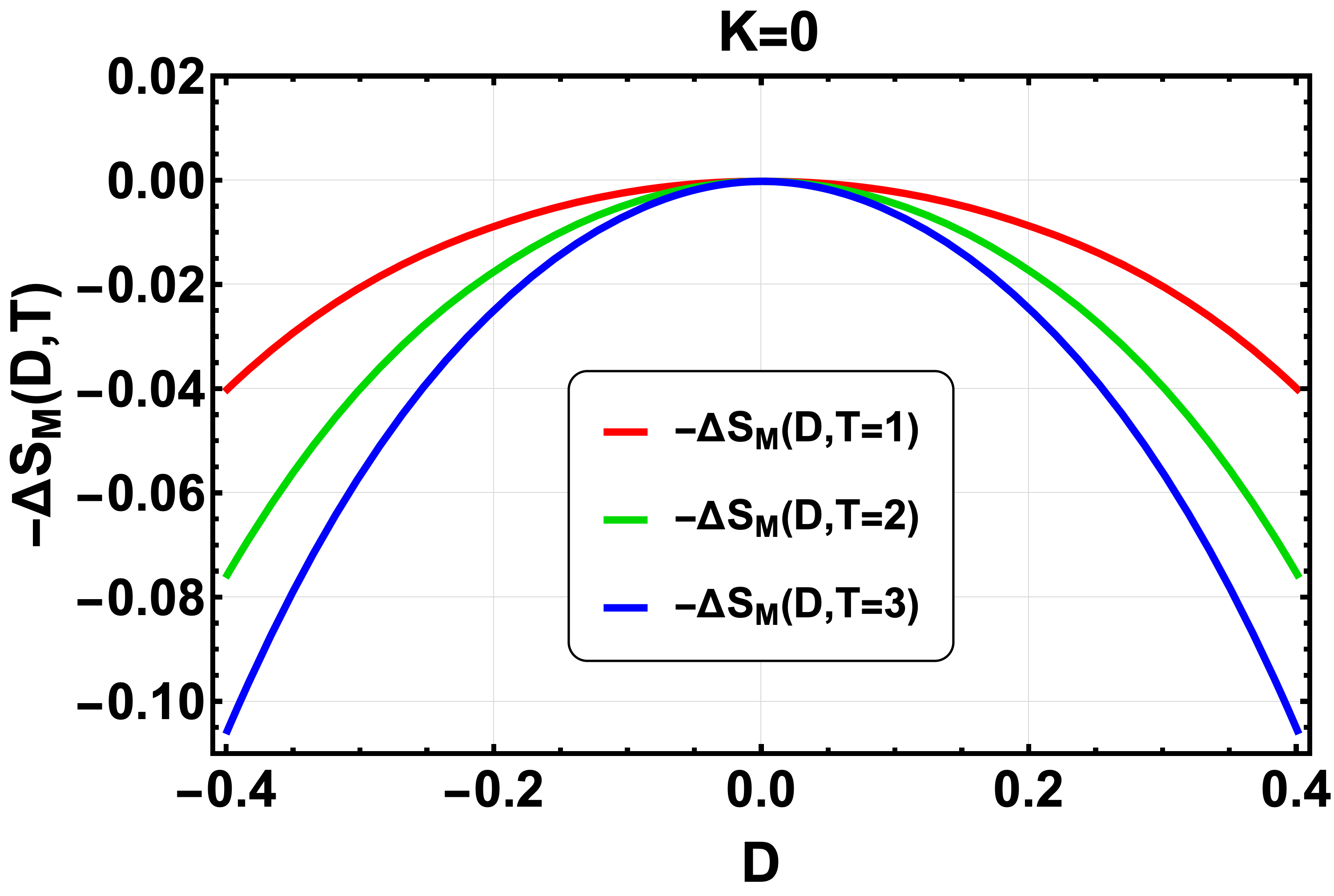}
    \caption{Isothermal entropy variation \(-\Delta S_M(D,T)= S_M(0,T) - S_M(D,T)\) for several temperatures and \(\,K=0\). The curves exhibit perfect symmetry under \(D\to -D\), reflecting the invariance of the magnonic spectrum with respect to the sign of the second–neighbor DM interaction. Since the caloric response is proportional to \(-\Delta S_M\), the system cools upon increasing \(D\).}
    \label{fig:deltaS_D}
\end{figure}

Fig.~\ref{fig:deltaS_D} illustrates the isothermal entropy change at $K=0$, defined as $-\Delta S_M(D, T) = S_M(0, T) - S_M(D, T)$, where the zero-DM configuration serves as the reference state. All curves display a pronounced and strictly even dependence on $D$, consistent with the spectral invariance under the transformation $D \to -D$ discussed above. For all temperatures, $-\Delta S_M(D,T) < 0$, indicating that a finite DM coupling increases the entropy relative to $D=0$. Since the caloric response is governed by $-\Delta S_M$, an isothermal increase of $D$ yields $-\Delta S_M < 0$, corresponding to a cooling effect: the system releases heat to the environment as $D$ increases. The magnitude of this response grows with temperature, reflecting the enhanced sensitivity of the thermal magnon population to DM-induced modifications of the hopping phases.

\begin{figure}[t]
    \centering
    \includegraphics[width=0.95\linewidth]{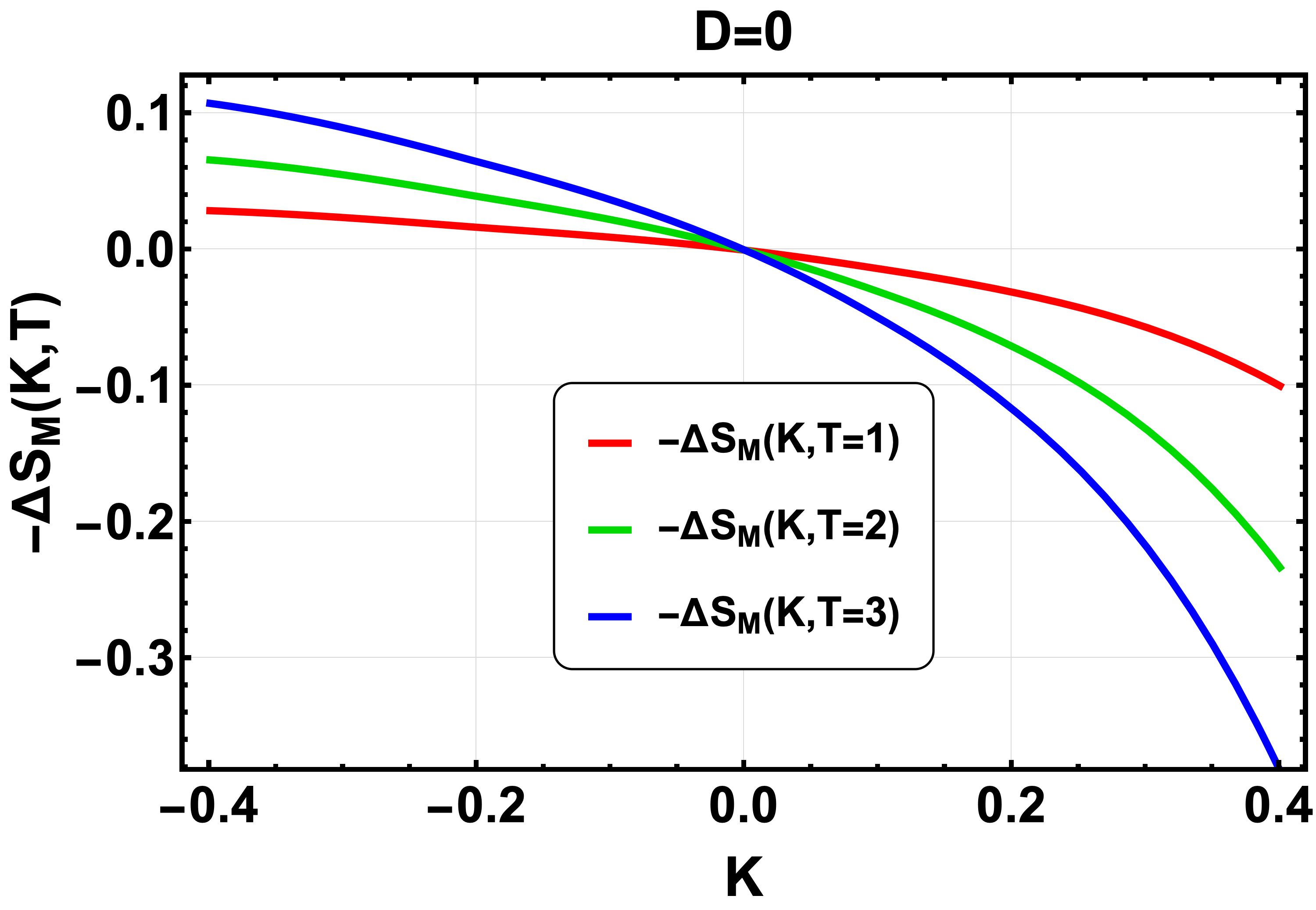}
    \caption{Isothermal entropy variation \(-\Delta S_M(K,T)=S_M(0,T)-S_M(K,T)\) for several temperatures and \( \,D=0\). The curves exhibit a strong asymmetry under \(K\to -K\). Negative \(K\) produces a direct caloric effect (\(-\Delta S_M>0\)), while positive \(K\) yields an inverse caloric effect (\(-\Delta S_M<0\)), reflecting the distinct thermal responses induced by the bond–anisotropic Kitaev interaction.}
    \label{fig:deltaS_K}
\end{figure}

Fig.~\ref{fig:deltaS_K} illustrates the isothermal entropy variation $-\Delta S_M(K,T) = S_M(K,T) - S_M(0,T)$ for several temperatures, where the $K=0$ state serves as the reference configuration. In contrast to the DM-driven response, the curves display a pronounced asymmetry with respect to \(K=0\), reflecting the non-symmetric modification of the magnonic spectrum induced by the bond-dependent Kitaev interaction. For negative values of \(K\), the entropy variation is positive, implying that \(-\Delta S_M>0\); thus, an isothermal increase of the Kitaev coupling in this regime leads to a direct caloric response in which the system absorbs heat, and its temperature rises. Conversely, for positive \(K\) the entropy variation becomes negative, and therefore \(-\Delta S_M<0\), corresponding to an inverse caloric response where the system cools as \(K\) increases. 

This highlights that the even parity of the DM response constrains the heat engine to a single operational regime, regardless of the sign of the coupling. In contrast, the Kitaev interaction, by generating distinct caloric responses for $K>0$ and $K<0$, allows the same material to access different operational regimes through the simple reversal of the exchange coupling.


\section{Stirling Engine}

In this section, we analyze the performance of the Stirling cycle using the microscopic couplings of the magnonic Hamiltonian as control parameters. Specifically, for the DM-driven cycle, we fix the upper coupling value at $D_H = 0.4$ and vary the lower value $D_L$ while maintaining $K = 0$. Similarly, for the Kitaev-driven cycle, we fix $K_H = 0.4$ and vary $K_L$ with $D = 0$. In both configurations, the efficiency is evaluated for various temperature pairs $(T_H, T_L)$ selected to ensure the system operates in the heat engine regime.

\begin{figure}[t]
    \centering
    \includegraphics[width=0.95\linewidth]{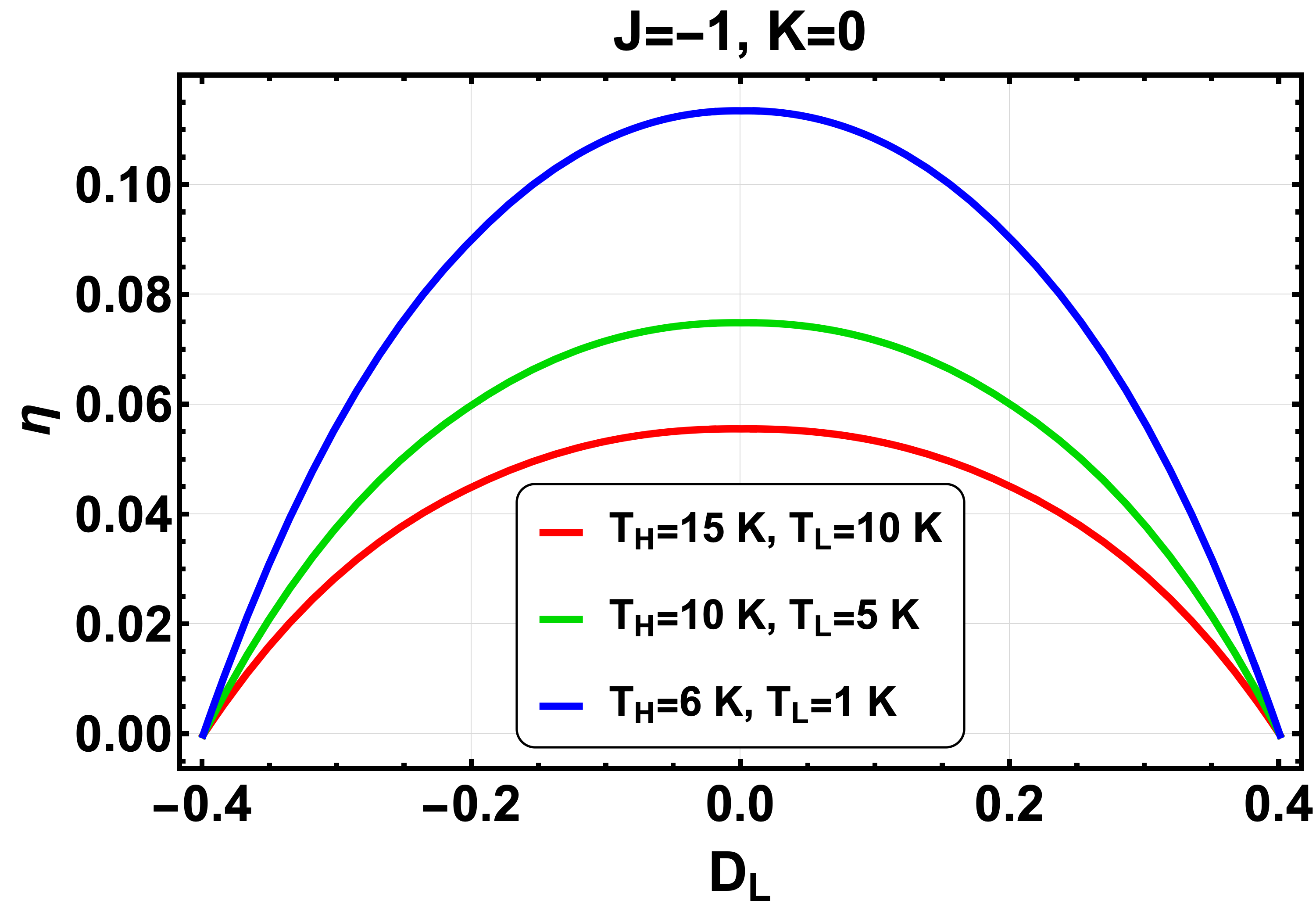}
    \caption{Efficiency $\eta$ of the Stirling Engine as a function of the lower $D_L$ for different temperature pairs $(T_H,T_L)$, $K=0$, and fixed upper value $D_H=0.4$. The curves show a symmetric maximum at $D_L=0$, reflecting the even dependence of the DM interaction.}
    \label{fig:eta_D}
\end{figure}

Fig.~\ref{fig:eta_D} presents the efficiency $\eta$ of the Stirling engine for the protocol in which the DM coupling serves as the control parameter. Here, the upper value is fixed at $D_H=0.4$, while the lower value $D_L$ is varied at $K=0$. Across all temperature pairs, the efficiency exhibits a symmetric concave profile with a well-defined maximum at $D_L=0$, decreasing monotonically as $|D_L|$ increases. This central peak stems from the parity of the interaction; specifically, it reflects the even dependence of the magnonic spectrum on the DM coupling, which yields identical thermodynamic responses for $\pm D$. While an increasing temperature gradient $T_H - T_L$ enhances the overall efficiency magnitude, the optimal operating point remains anchored at $D_L=0$, where the working medium achieves the most favorable balance between entropy change.
 
\begin{figure}[h!]
    \centering
    \includegraphics[width=0.95\linewidth]{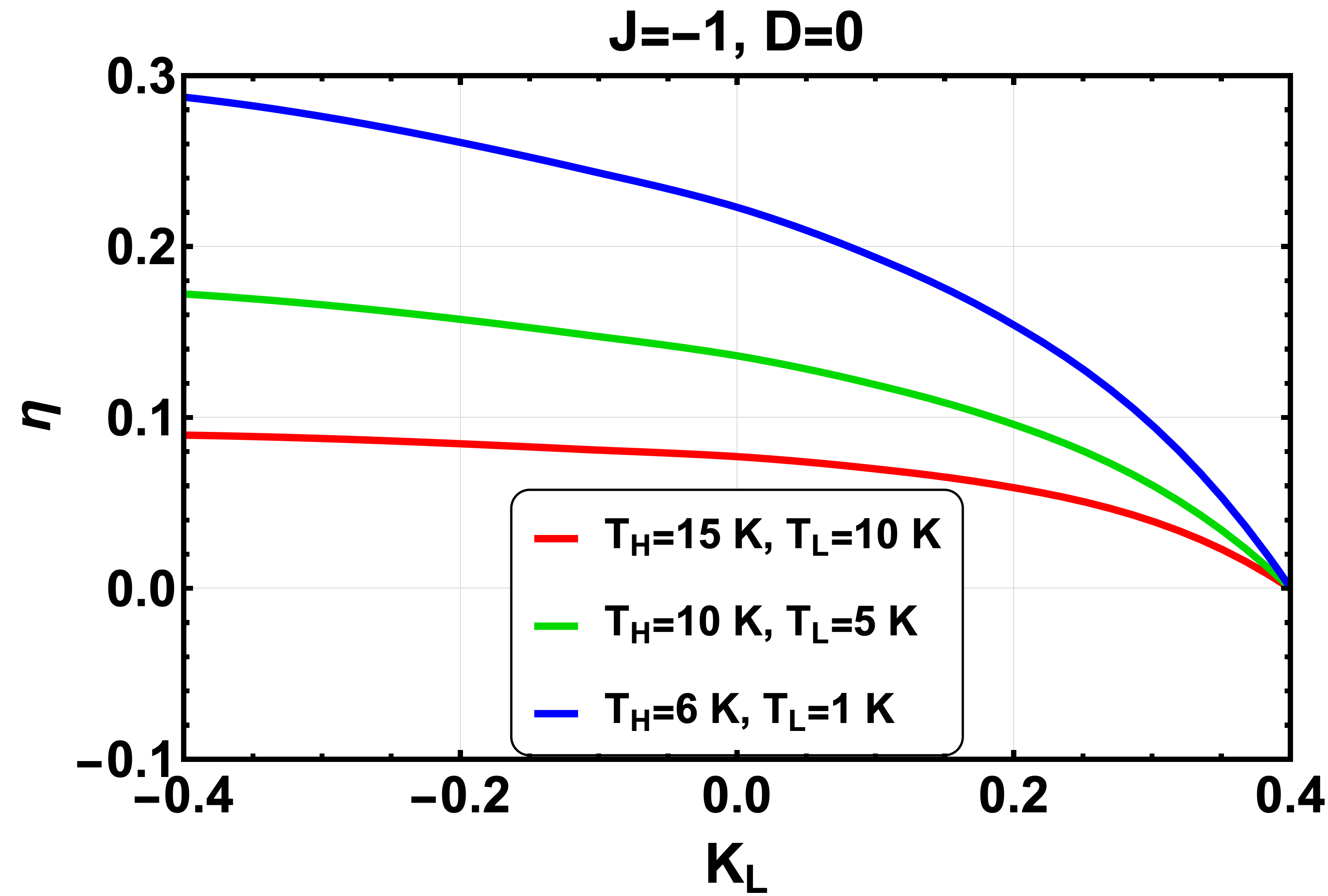}
    \caption{Efficiency $\eta$ of the Stirling Engine as a function of the lower Kitaev coupling $K_L$ for different temperature pairs $(T_H,T_L)$, with $D=0$, and fixed upper value $K_H=0.4$. The curves exhibit a strong asymmetry under $K_L\to -K_L$.}
    \label{fig:eta_K}
\end{figure}

Fig.~\ref{fig:eta_K} illustrates the efficiency of the Stirling cycle when the Kitaev exchange serves as the control parameter, with $K_H=0.4$ fixed, $D=0$, and $K_L$ as the variable. In stark contrast to the DM-driven case, the efficiency curves are strongly asymmetric with respect to $K_L=0$. This behavior stems from the uneven redistribution of the magnonic spectral weight under the transformation $K \to -K$, leading to a markedly different efficiency landscape.
A direct comparison with the DM-driven protocol reveals two fundamental differences in engine performance. First, the parity symmetry observed in the DM case is completely lifted: while the DM-driven efficiency follows a symmetric parabolic profile that vanishes at $D_L = \pm D_H$, the Kitaev-controlled engine exhibits a stark directional dependence, attaining its maximum efficiency at $K_L = -0.4$. As $K_L$ becomes increasingly negative, the system approaches a saturation regime, a feature that underscores the intensive spectral weight redistribution and magnonic gap modulation inherent to bond-dependent anisotropy.

Second, under identical thermal reservoirs, the efficiency achieved via Kitaev modulation significantly exceeds that of its DM-driven counterpart. This performance enhancement originates from the more pronounced distortion of the magnonic dispersion induced by the Kitaev exchange. Such a drastic spectral reshaping enables the working medium to undergo substantially larger variations in entropy during both the isothermal and isoparametric strokes. Consequently, the Kitaev-mediated Stirling cycle demonstrates a superior energy-conversion capability and greater operational versatility for nanoscale heat management.

\section{Experimental feasibility and control of $K$ and $D$}
\label{sec:exp_feasibility}
The Stirling cycles analyzed in this work rely on the quasistatic modulation of effective spin--exchange couplings, rather than on mechanical coordinates in the usual sense.
From an experimental perspective, the Kitaev exchange $K$ and the Dzyaloshinskii--Moriya interaction $D$ should be understood as effective low--energy parameters of the magnon Hamiltonian, whose values can be tuned indirectly through external fields, lattice distortions, and interface or
heterostructure engineering. Materials with strong spin--orbit coupling on honeycomb or related lattices provide a natural platform where both Kitaev--type anisotropic exchange and Dzyaloshinskii--Moriya interactions can become relevant at low energies. Prominent examples include $\alpha$--RuCl$_3$ and other layered van der Waals magnets, where anisotropic exchange interactions have been extensively discussed \cite{Jackeli2009,Winter2017,Janssen2019}. In these systems, an external magnetic field can stabilize a field--polarized regime with well--defined magnon excitations, which provides a suitable setting for magnonic thermodynamic protocols \cite{Banerjee2018,suzuki2021proximate,ponomaryov2020nature,zhou2023possible}.

The Kitaev exchange $K$ is particularly sensitive to lattice geometry and
bonding angles.
As a consequence, its effective strength can be modified by hydrostatic
pressure, uniaxial strain, or substrate--induced distortions, which alter the
overlap between spin--orbit--entangled orbitals and hence the anisotropic
superexchange pathways \cite{yadav2018strain,kim2013strain,cenker2022reversible}.
Related strategies include interface engineering in thin films or
heterostructures, where local symmetry breaking and structural relaxation can
selectively enhance or suppress bond--dependent interactions
\cite{miao2021straintronics,du2021strain,li2020recent}.
Importantly, the Stirling cycle proposed here requires relative variations of
$K$ within a finite window, rather than absolute control over its magnitude.

The Dzyaloshinskii--Moriya interaction $D$, on the other hand, originates from
the combination of spin--orbit coupling and broken inversion symmetry.
Its magnitude and orientation can therefore be tuned by controlling structural
asymmetry, for instance through interfaces, surfaces, gating, or stacking order
in layered systems \cite{Fert2013,Tokura2014,Soumyanarayanan2017}.
In addition, several works have shown that electric fields and strain--mediated
magnetoelectric routes can modulate interfacial DMI in insulating magnetic
heterostructures \cite{Katsura2005,lei2013strain,zhou2022piezoelectric}.
These mechanisms provide plausible routes to implement the quasistatic
modulation of $D$ required by DM--driven protocols. Therefore, the $(D,K)$ parameter space may be experimentally accessed through a combination of electric field control and strain or pressure engineering. Throughout this work, the external magnetic field $B$ is kept fixed during the
cycle. Experimentally, this corresponds to operating within a stable field--polarized phase, where the magnon description remains valid and the low--energy spectrum is well separated from instabilities toward competing magnetic orders. Such conditions are routinely explored in candidate Kitaev materials under laboratory magnetic fields \cite{Banerjee2018,ponomaryov2020nature,ozel2019magnetic,Kasahara2018}.

Finally, we emphasize that the present analysis is formulated at the level of
equilibrium thermodynamics and does not rely on a specific dynamical protocol.
The Stirling cycle should therefore be viewed as a conceptual framework that
demonstrates how coupling--controlled caloric responses of bosonic
quasiparticles can be harvested to produce work.
The control mechanisms discussed above indicate that the required modulation of
$K$ and $D$ is physically plausible in existing or near--future experimental
platforms, opening the door to magnonic heat engines driven by exchange interactions.

\section{Conclusion}
In summary, we have investigated the thermodynamic properties and the performance of a magnonic Stirling heat engine governed by a Heisenberg--Kitaev Hamiltonian with second-neighbor Dzyaloshinskii--Moriya (DM) interactions. By employing linear spin-wave theory, we established a rigorous connection between the symmetries of the microscopic exchange couplings and the macroscopic caloric responses of the bosonic working medium. We demonstrated that the DM interaction, which contributes only a complex phase to the hopping amplitudes, modifies the magnon spectrum quadratically. This mechanism preserves the spectral symmetry under $D \to -D$, restricting the thermodynamic observables and the engine's efficiency to strictly even profiles, thereby limiting the system to a single operational regime.

In stark contrast, the bond-dependent Kitaev exchange introduces a pronounced asymmetry under $K \to -K$, fundamentally reshaping the low-energy spectral weight and the magnonic gap. This spectral distortion not only enables both direct and inverse caloric effects depending on the sign of $K$, but also yields a significantly enhanced Stirling cycle efficiency compared to the DM-driven protocol. Furthermore, the Kitaev-mediated engine exhibits a high-efficiency saturation regime at negative coupling values, highlighting its superior capability for energy conversion.

Given the experimental tunability of these effective exchange parameters through strain engineering, electric fields, and heterostructure design in van der Waals magnets and iridate-based compounds, our findings suggest that Kitaev-active materials are highly versatile and efficient platforms for nanoscale heat management and solid-state quantum thermodynamics.

\section*{Acknowledgments}
B.C., M.H.G., F.J.P., and P.V. acknowledge support from FONDECYT (Chile) under Grants No.~1250173 and No.~1240582. R.E.T. and N.V.-S. acknowledge funding from FONDECYT Regular Grants No.~1230747 and No.~1250364, respectively. F.B. acknowledges funding from FONDECYT Regular Grant. 1231210. P.V. acknowledges support from CEDENNA Grant No.~250002. B.C. and M.H.G. acknowledge support from PUCV and the Direcci\'on de Postgrado of UTFSM. B.C. further acknowledges support from the Programa de Incentivo a la Iniciaci\'on Cient\'ifica (PIIC) No.~004 and ANID Becas/Doctorado Nacional 21250015.

\appendix
\section{Bulk magnon spectrum} \label{Bulk_magnon}

In the main text, the quadratic magnon Hamiltonian is written as
\begin{equation}
H_m=\frac{S}{2}\sum_{\bm{k}}
\Psi_{\bm{k}}^\dagger M_{\bm{k}} \Psi_{\bm{k}},
\qquad
\Psi_{\bm{k}}=(a_{\bm{k}},\,b_{\bm{k}},\,a^\dagger_{-\bm{k}},\,b^\dagger_{-\bm{k}})^T ,
\end{equation}
where the bosonic Bogoliubov--de Gennes (BdG) matrix reads
\begin{equation}
M_{\bm{k}}=
\begin{pmatrix}
A_{\bm{k}} & B_{\bm{k}}\\
B_{\bm{k}}^\dagger & A_{\bm{k}}
\end{pmatrix}.
\end{equation}
The $2\times2$ blocks are given by
\begin{equation}
A_{\bm{k}}=
\begin{pmatrix}
\kappa_0+\kappa_{1,\bm{k}} & \kappa_{2,-\bm{k}}\\
\kappa_{2,\bm{k}} & \kappa_0-\kappa_{1,\bm{k}}
\end{pmatrix},
\qquad
B_{\bm{k}}=
\begin{pmatrix}
0 & \kappa_{3,-\bm{k}}\\
\kappa_{3,\bm{k}} & 0
\end{pmatrix},
\end{equation}
with $\kappa_0$, $\kappa_{1,\bm{k}}$, $\kappa_{2,\bm{k}}$ and $\kappa_{3,\bm{k}}$ defined in Eqs.~(5)--(8) of the main text.

The magnon energies are obtained from the generalized bosonic eigenvalue problem
\begin{equation}
\eta M_{\bm{k}} w_{\bm{k}}=\varepsilon_{\bm{k}} w_{\bm{k}},
\qquad
\eta=\mathrm{diag}(1,1,-1,-1),
\label{eq:Bosonicproblem}
\end{equation}
where $w_{\bm{k}}=(u_{\bm{k}},v_{\bm{k}})^T$ with $u_{\bm{k}},v_{\bm{k}}\in\mathbb{C}^2$. As stated in the main text, we restrict to the case $\Gamma^z=0$. In this limit, the anomalous term $\kappa_{3,\bm{k}}\kappa_{3,-\bm{k}}$ becomes real up to a phase factor, allowing us to write $\kappa_{3,\bm{k}}\kappa_{3,-\bm{k}}=\vert \kappa_{3,\bm{k}}\vert^2$. Thus, eliminating $v_{\bm k}$ in Eq. \eqref{eq:Bosonicproblem} yields a reduced equation for $u_{\bm k}$,
\begin{equation}
Q_{\bm k}\,u_{\bm k}=\varepsilon_{\bm k}^2 u_{\bm k},
\qquad
Q_{\bm k}=(A_{\bm k}-B_{\bm k})(A_{\bm k}+B_{\bm k}).
\end{equation}

which reduces the original $4\times4$ BdG problem to a $2\times2$ matrix $Q_{\bm{k}}$. Since $Q_{\bm{k}}$ is a $2\times2$ matrix, its eigenvalues are obtained analytically,
\begin{equation}
\varepsilon_{\bm{k},\pm}^2=
\frac{\mathrm{Tr}\,Q_{\bm{k}}}{2}
\pm
\frac{1}{2}
\sqrt{(\mathrm{Tr}\,Q_{\bm{k}})^2-4\det Q_{\bm{k}}}.
\end{equation}
Evaluating the trace and determinant yields
\begin{align}
\varepsilon_{\bm{k},\pm}^{2}=\kappa_0^2+\kappa_{1,\bm{k}}^2+\vert\kappa_{2,\bm{k}}\vert^2-\vert\kappa_{3,\bm{k}}\vert^2\pm\sqrt{\Delta_{\bm{k}}},
\label{eqAp:magnondispersion}
\end{align}
with
\begin{align} \nonumber\Delta_{\bm{k}}&=4\kappa_0^2\kappa_{1,\bm{k}}^2+4\kappa_0^2\vert \kappa_{2,\bm{k}}\vert^2-4\kappa_{1,\bm{k}}^2\vert\kappa_{3,\bm{k}}\vert^2\\
    &+\left(\kappa_{2,\bm{k}}\kappa_{3,-\bm{k}}-\kappa_{2,-\bm{k}}\kappa_{3,\bm{k}}\right)^2.
    \label{eqAp:determ}
\end{align}
The physical magnon branches correspond to the positive square roots
$\varepsilon_{\bm{k}}^{\pm}=\sqrt{\epsilon_{\bm{k},\pm}^{2}}$. For completeness, if $\Gamma_z\neq0$ the coefficient
$\kappa_{3,\bm{k}}=i\Gamma_z e^{-i\bm{k}\cdot\delta_z}
+K_x e^{-i\bm{k}\cdot\delta_x}-K_y e^{-i\bm{k}\cdot\delta_y}$
acquires an explicit imaginary contribution.
In that case $\kappa_{3,-\bm{k}}\neq\kappa_{3,\bm{k}}^{*}$ in general, and the
simplification $\kappa_{3,\bm{k}}\kappa_{3,-\bm{k}}=|\kappa_{3,\bm{k}}|^{2}$
no longer holds. Consequently, the analytical reduction used above is
not guaranteed. The magnon spectrum must then be obtained from the
full bosonic BdG eigenvalue problem, i.e., by performing a paraunitary
diagonalization of the $4\times4$ bosonic BdG Hamiltonian.

\bibliography{biblio}

@article{Janssen2019,
  title = {Heisenberg–Kitaev physics in magnetic fields},
  volume = {31},
  ISSN = {1361-648X},
  url = {http://dx.doi.org/10.1088/1361-648X/ab283e},
  DOI = {10.1088/1361-648x/ab283e},
  number = {42},
  journal = {Journal of Physics: Condensed Matter},
  publisher = {IOP Publishing},
  author = {Janssen,  Lukas and Vojta,  Matthias},
  year = {2019},
  month = jul,
  pages = {423002}
}

@article{miao2021straintronics,
  title={Straintronics with van der Waals materials},
  author={Miao, Feng and Liang, Shi-Jun and Cheng, Bin},
  journal={npj Quantum Materials},
  volume={6},
  number={1},
  pages={59},
  year={2021},
  publisher={Nature Publishing Group UK London}
}

@article{joshi2018topological,
  title={Topological excitations in the ferromagnetic Kitaev-Heisenberg model},
  author={Joshi, Darshan G},
  journal={Physical Review B},
  volume={98},
  number={6},
  pages={060405},
  year={2018},
  publisher={APS}
}

@article{mcclarty2018topological,
  title={Topological magnons in Kitaev magnets at high fields},
  author={McClarty, PA and Dong, X-Y and Gohlke, M and Rau, JG and Pollmann, F and Moessner, R and Penc, K},
  journal={Physical Review B},
  volume={98},
  number={6},
  pages={060404},
  year={2018},
  publisher={APS}
}

@article{vidal2024magnonic,
  title={Magnonic Otto thermal machine},
  author={Vidal-Silva, Nicolas and Pe{\~n}a, Francisco J and Troncoso, Roberto E and Vargas, Patricio},
  journal={Physical Review Research},
  volume={6},
  number={3},
  pages={033164},
  year={2024},
  publisher={APS}
}

@article{du2021strain,
  title={Strain engineering in 2D material-based flexible optoelectronics},
  author={Du, Junli and Yu, Huihui and Liu, Baishan and Hong, Mengyu and Liao, Qingliang and Zhang, Zheng and Zhang, Yue},
  journal={Small Methods},
  volume={5},
  number={1},
  pages={2000919},
  year={2021},
  publisher={Wiley Online Library}
}

@article{li2020recent,
  title={Recent advances in strain-induced piezoelectric and piezoresistive effect-engineered 2D semiconductors for adaptive electronics and optoelectronics},
  author={Li, Feng and Shen, Tao and Wang, Cong and Zhang, Yupeng and Qi, Junjie and Zhang, Han},
  journal={Nano-Micro Letters},
  volume={12},
  pages={1--44},
  year={2020},
  publisher={Springer}
}

@article{cenker2022reversible,
  title={Reversible strain-induced magnetic phase transition in a van der Waals magnet},
  author={Cenker, John and Sivakumar, Shivesh and Xie, Kaichen and Miller, Aaron and Thijssen, Pearl and Liu, Zhaoyu and Dismukes, Avalon and Fonseca, Jordan and Anderson, Eric and Zhu, Xiaoyang and others},
  journal={Nature Nanotechnology},
  volume={17},
  number={3},
  pages={256--261},
  year={2022},
  publisher={Nature Publishing Group UK London}
}

@article{lei2013strain,
  title={Strain-controlled magnetic domain wall propagation in hybrid piezoelectric/ferromagnetic structures},
  author={Lei, Na and Devolder, Thibaut and Agnus, Guillaume and Aubert, Pascal and Daniel, Laurent and Kim, Joo-Von and Zhao, Weisheng and Trypiniotis, Theodossis and Cowburn, Russell P and Chappert, Claude and others},
  journal={Nature communications},
  volume={4},
  number={1},
  pages={1378},
  year={2013},
  publisher={Nature Publishing Group UK London}
}

@article{colpa1978diagonalization,
  title={Diagonalization of the quadratic boson hamiltonian},
  author={Colpa, JHP},
  journal={Physica A: Statistical Mechanics and its Applications},
  volume={93},
  number={3-4},
  pages={327--353},
  year={1978},
  publisher={Elsevier}
}

@article{ozel2019magnetic,
  title={Magnetic field-dependent low-energy magnon dynamics in $\alpha$--RuCl 3},
  author={Ozel, Ilkem Ozge and Belvin, Carina A and Baldini, Edoardo and Kimchi, Itamar and Do, Seunghwan and Choi, Kwang-Yong and Gedik, Nuh},
  journal={Physical Review B},
  volume={100},
  number={8},
  pages={085108},
  year={2019},
  publisher={APS}
}

@article{flebus20242024,
  title={The 2024 magnonics roadmap},
  author={Flebus, Benedetta and Grundler, Dirk and Rana, Bivas and Otani, Yoshichika and Barsukov, Igor and Barman, Anjan and Gubbiotti, Gianluca and Landeros, Pedro and Akerman, Johan and Ebels, Ursula and others},
  journal={Journal of Physics: Condensed Matter},
  volume={36},
  number={36},
  pages={363501},
  year={2024},
  publisher={IOP Publishing}
}

@article{chumak2015magnon,
  title={Magnon spintronics},
  author={Chumak, Andrii V and Vasyuchka, Vitaliy I and Serga, Alexander A and Hillebrands, Burkard},
  journal={Nature physics},
  volume={11},
  number={6},
  pages={453--461},
  year={2015},
  publisher={Nature Publishing Group UK London}
}

@article{yuan2022quantum,
  title={Quantum magnonics: When magnon spintronics meets quantum information science},
  author={Yuan, HY and Cao, Yunshan and Kamra, Akashdeep and Duine, Rembert A and Yan, Peng},
  journal={Physics Reports},
  volume={965},
  pages={1--74},
  year={2022},
  publisher={Elsevier}
}

@article{wei2024strain,
  title={Strain-engineered magnon states in two-dimensional ferromagnetic monolayers},
  author={Wei, Bin and Zhu, Jia-Ji and Song, Yun and Chang, Kai},
  journal={Physical Review Research},
  volume={6},
  number={1},
  pages={013210},
  year={2024},
  publisher={APS}
}

@article{zhou2022piezoelectric,
  title={Piezoelectric strain-controlled magnon spin current transport in an antiferromagnet},
  author={Zhou, Yongjian and Guo, Tingwen and Qiao, Leilei and Wang, Qian and Zhu, Meng and Zhang, Jia and Liu, Quan and Zhao, Mingkun and Wan, Caihua and He, Wenqing and others},
  journal={Nano Letters},
  volume={22},
  number={12},
  pages={4646--4653},
  year={2022},
  publisher={ACS Publications}
}

@article{yadav2018strain,
  title={Strain-and pressure-tuned magnetic interactions in honeycomb Kitaev materials},
  author={Yadav, Ravi and Rachel, Stephan and Hozoi, Liviu and van den Brink, Jeroen and Jackeli, George},
  journal={Physical Review B},
  volume={98},
  number={12},
  pages={121107},
  year={2018},
  publisher={APS}
}

@article{kim2013strain,
  title={Strain-induced topological insulator phase and effective magnetic interactions in Li 2 IrO 3},
  author={Kim, Heung-Sik and Kim, Choong H and Jeong, Hogyun and Jin, Hosub and Yu, Jaejun},
  journal={Physical Review B—Condensed Matter and Materials Physics},
  volume={87},
  number={16},
  pages={165117},
  year={2013},
  publisher={APS}
}

@article{holstein1940field,
  title={Field dependence of the intrinsic domain magnetization of a ferromagnet},
  author={Holstein, Theodore and Primakoff, Henry},
  journal={Physical Review},
  volume={58},
  number={12},
  pages={1098},
  year={1940},
  publisher={APS}
}

@article{suzuki2021proximate,
  title={Proximate ferromagnetic state in the Kitaev model material $\alpha$-RuCl3},
  author={Suzuki, Hakuto and Liu, Huimei and Bertinshaw, Joel and Ueda, Kentaro and Kim, H and Laha, Sourav and Weber, Daniel and Yang, Zichen and Wang, Lichen and Takahashi, Hiroto and others},
  journal={Nature communications},
  volume={12},
  number={1},
  pages={4512},
  year={2021},
  publisher={Nature Publishing Group UK London}
}

@article{zhou2023possible,
  title={Possible intermediate quantum spin liquid phase in $\alpha$-RuCl3 under high magnetic fields up to 100 T},
  author={Zhou, Xu-Guang and Li, Han and Matsuda, Yasuhiro H and Matsuo, Akira and Li, Wei and Kurita, Nobuyuki and Su, Gang and Kindo, Koichi and Tanaka, Hidekazu},
  journal={Nature Communications},
  volume={14},
  number={1},
  pages={5613},
  year={2023},
  publisher={Nature Publishing Group UK London}
}

@article{ponomaryov2020nature,
  title={Nature of magnetic excitations in the high-field phase of $\alpha$-RuCl 3},
  author={Ponomaryov, AN and Zviagina, Larysa and Wosnitza, Joachim and Lampen-Kelley, P and Banerjee, A and Yan, J-Q and Bridges, CA and Mandrus, DG and Nagler, SE and Zvyagin, SA},
  journal={Physical Review Letters},
  volume={125},
  number={3},
  pages={037202},
  year={2020},
  publisher={APS}
}

@article{Jackeli2009,
  author = {G. Jackeli and G. Khaliullin},
  title = {Mott Insulators in the Strong Spin-Orbit Coupling Limit: From Heisenberg to a Quantum Compass and Kitaev Models},
  journal = {Phys. Rev. Lett.},
  volume = {102},
  pages = {017205},
  year = {2009}
}

@article{Winter2017,
  author = {S. M. Winter et al.},
  journal = {J. Phys.: Condens. Matter},
  volume = {29},
  pages = {493002},
  year = {2017}
}

@article{Banerjee2018,
  author = {A. Banerjee et al.},
  journal = {npj Quantum Materials},
  volume = {3},
  pages = {8},
  year = {2018}
}

@article{Fert2013,
  author = {A. Fert et al.},
  journal = {Nat. Nanotechnol.},
  volume = {8},
  pages = {152},
  year = {2013}
}

@article{Tokura2014,
  author = {Y. Tokura and N. Kanazawa},
  journal = {Chem. Rev.},
  volume = {114},
  pages = {7537},
  year = {2014}
}

@article{Katsura2005,
  title   = {Spin Current and Magnetoelectric Effect in Noncollinear Magnets},
  author  = {Katsura, H. and Nagaosa, N. and Balatsky, A. V.},
  journal = {Phys. Rev. Lett.},
  volume  = {95},
  pages   = {057205},
  year    = {2005},
  doi     = {10.1103/PhysRevLett.95.057205}
}

@article{Soumyanarayanan2017,
  author = {A. Soumyanarayanan et al.},
  journal = {Nat. Mater.},
  volume = {16},
  pages = {898},
  year = {2017}
}

@article{Kasahara2018,
  title   = {Unusual Thermal Hall Effect in a Kitaev Spin Liquid Candidate $\alpha$-RuCl$_3$},
  author  = {Kasahara, Y. and Sugii, K. and Ohnishi, T. and Shiozaki, R. and Matsumoto, Y. and Botana, A. S. and Yoshitake, J. and Nasu, J. and Motome, Y. and Takagi, H. and Matsuda, Y.},
  journal = {Phys. Rev. Lett.},
  volume  = {120},
  pages   = {217205},
  year    = {2018},
  doi     = {10.1103/PhysRevLett.120.217205}
}

@article{Pervez2026,
  title={Quantum Otto heat-engine with Kitaev-Heisenberg cluster:
  Possible roles of frustration, magnons, and duality},
  author={Pervez, Sheikh Moonsun and Mandal, Saptarshi},
  journal={arXiv preprint arXiv:2601.03826},
  year={2026}
}

@article{demokritov2006bose,
  title={Bose--Einstein condensation of quasi-equilibrium magnons at room temperature under pumping},
  author={Demokritov, Sergej O and Demidov, Vladislav E and Dzyapko, Oleksandr and Melkov, Gennadii A and Serga, Alexandar A and Hillebrands, Burkard and Slavin, Andrei N},
  journal={Nature},
  volume={443},
  number={7110},
  pages={430--433},
  year={2006},
  publisher={Nature Publishing Group UK London}
}

@article{cornelissen2016magnon,
  title={Magnon spin transport driven by the magnon chemical potential in a magnetic insulator},
  author={Cornelissen, Ludo J and Peters, Kevin JH and Bauer, Gerrit EW and Duine, Rembert A and van Wees, Bart J},
  journal={Physical Review B},
  volume={94},
  number={1},
  pages={014412},
  year={2016},
  publisher={APS}
}

@article{Feldmann2003,
  author = {Feldmann, T. and Kosloff, R.},
  title = {Quantum four-stroke heat engine: Thermodynamic observables in a model with intrinsic friction},
  journal = {Phys. Rev. E},
  volume = {68},
  pages = {016101},
  year = {2003},
  doi = {10.1103/PhysRevE.68.016101}
}

@article{Quan2007,
  author = {Quan, H. T. and Liu, Yu-xi and Sun, C. P. and Nori, F.},
  title = {Quantum thermodynamic cycles and quantum heat engines},
  journal = {Phys. Rev. E},
  volume = {76},
  pages = {031105},
  year = {2007},
  doi = {10.1103/PhysRevE.76.031105}
}

@article{Thomas2011,
  author = {Thomas, G. and Johal, R. S.},
  title = {Coupled quantum Otto cycle},
  journal = {Phys. Rev. E},
  volume = {83},
  pages = {031135},
  year = {2011},
  doi = {10.1103/PhysRevE.83.031135}
}

@article{Campisi2016,
  author = {Campisi, M. and Fazio, R.},
  title = {The power of a critical heat engine},
  journal = {Nature Communications},
  volume = {7},
  pages = {11895},
  year = {2016},
  doi = {10.1038/ncomms11895}
}

@article{Kieu2004,
  author = {Kieu, T. D.},
  title = {The Second Law, Maxwell’s Demon, and Work Derivable from Quantum Heat Engines},
  journal = {Phys. Rev. Lett.},
  volume = {93},
  pages = {140403},
  year = {2004},
  doi = {10.1103/PhysRevLett.93.140403}
}

@article{Dillenschneider2009,
  author = {Dillenschneider, R. and Lutz, E.},
  title = {Energetics of quantum correlations},
  journal = {Europhys. Lett.},
  volume = {88},
  pages = {50003},
  year = {2009},
  doi = {10.1209/0295-5075/88/50003}
}

@article{YungerHalpern2019,
  author = {Yunger Halpern, N. and Faist, P. and Oppenheim, J. and Winter, A.},
  title = {Microcanonical and resource-theoretic derivations of the thermal state of a quantum system with noncommuting charges},
  journal = {Phys. Rev. E},
  volume = {99},
  pages = {032131},
  year = {2019},
  doi = {10.1103/PhysRevE.99.032131}
}

@article{Shindou2013,
  author = {Shindou, R. and Matsumoto, R. and Murakami, S. and Ohe, J.},
  title = {Topological chiral magnonic edge mode in a magnonic crystal},
  journal = {Phys. Rev. B},
  volume = {87},
  pages = {174427},
  year = {2013},
  doi = {10.1103/PhysRevB.87.174427}
}

@article{Matsumoto2011,
  author = {Matsumoto, R. and Murakami, S.},
  title = {Theoretical prediction of a rotating magnon wave packet in ferromagnets},
  journal = {Phys. Rev. Lett.},
  volume = {106},
  pages = {197202},
  year = {2011},
  doi = {10.1103/PhysRevLett.106.197202}
}

@book{AltlandSimons,
  author = {Altland, A. and Simons, B.},
  title = {Condensed Matter Field Theory},
  publisher = {Cambridge University Press},
  year = {2010}
}

\end{document}